\documentclass[aps,prb,twocolumn,preprintnumbers]{revtex4-1}

\usepackage{graphicx}
\usepackage{dcolumn}
\usepackage{bm}
\usepackage{amsmath}
\usepackage{amssymb}
\usepackage{hyperref,xcolor,dsfont}
\renewcommand{\vec}[1]{{{\mathbf{\boldsymbol #1}}}}

\begin{document}
\title{Universality of annihilation barriers of large magnetic skyrmions in chiral and frustrated magnets}
\author{Benjamin Heil$^1$}
\author{Achim Rosch$^1$}
\author{Jan Masell$^{1,2}$}
\affiliation{
$^1$ Institute for Theoretical Physics, University of Cologne, 50937 Cologne, Germany\\
$^2$ RIKEN Center for Emergent Matter Science, Wako, Saitama 351-0198, Japan
}
\date{\today}

\begin{abstract}
Magnetic skyrmions are whirls in the magnetization whose topological winding number promises stability against thermal fluctuations and defects. 
They can only decay via singular spin configurations.
We analyze the corresponding energy barriers of skyrmions in a magnetic monolayer for two distinct stabilization mechanisms, i.e., Dzyaloshinskii-Moriya interaction (DMI) and competing interactions.
Based on our numerically calculated collapse paths on an atomic lattice, we derive analytic expressions for the saddle point textures and energy barriers of large skyrmions. 
The sign of the spin stiffness and the sign of 4th-order derivative terms in the classical field theory determines the nature of the saddle point and thus the height of the energy barrier. 
In the most common case for DMI-stabilized skyrmions, positive stiffness and negative 4th-order term, the saddle point energy approaches a universal upper limit described by an effective continuum theory. 
For skyrmions stabilized by frustrating interactions, the stiffness is negative and the energy barrier arises mainly from the core of a singular vortex configuration. 
\end{abstract}

%\pacs{75.78.-n,75.75.-c,12.39.Dc}
% Skyrmions:                        12.39.Dc
% Vortices in magnetic thin films:  75.70.Kw
% Vortex pinning (Superconductivity)74.25.Wx
% Vortex dynamics (fluid flow)      47.32.C-
% Dynamics of magnetization         75.78.-n
% magnetic properties of nanostructures 75.75.-c
% Magnetoelectronics/Spintronics: spin transport effects 75.76.+j
\maketitle

%----------------------------------------------------------------------------------------
%----------------------------------------------------------------------------------------
%----------------------------------------------------------------------------------------
\section{Introduction}
%----------------------------------------------------------------------------------------

A magnetic skyrmion is a magnetic texture where the spin-orientation varies smoothly over many lattice constants\cite{bogdanov1989thermodynamically,muhlbauer2009skyrmion,nagaosa2013topological,wiesendanger2016nanoscale,fert2017magnetic,everschorsitte2018perspective}. 
Its distinctive feature is the topological winding number: the magnetization of each skyrmion winds once in all possible directions. 
Arguably, the most fundamental property of a skyrmion is its stability: 
it is protected by an energy barrier from its destruction by thermal fluctuations\cite{bogdanov1999stability,buttner2018theory,wang2018atheory,bernandmantel2018theskyrmionbubble}.  
The size of this energy barrier is therefore an important quantity as it determines in what sense a skyrmion is "topologically protected". 
From the application perspective, the energy barrier determines not only the thermal stability of skyrmions but, for example, also how much energy is needed to create skyrmions.

Smooth deformations of the magnetic texture cannot change the topology, therefore the creation or destruction of a skyrmion can only occur via singular spin configurations, where the spin varies rapidly on the atomic scale\cite{nagaosa2013topological,milde2013unwinding}. 
A central question both of fundamental and practical importance is whether the energy barrier is ultimately determined by this singular spin configuration and, hence, the non-universal local physics on the length scale of a lattice constant, or whether it is governed by the universal physics encoded in (classical) continuum field theories.
If the first scenario holds, it is often very difficult to calculate the relevant energy barriers: 
most experimental systems are metallic\cite{nagaosa2013topological,everschorsitte2018perspective} and a full quantum mechanical calculation of an unstable high-energy state, necessarily including the electronic degree of freedom, would be very challenging. 
We will show that for a wide class of skyrmion systems the second scenario holds. 
Despite the singular nature of the skyrmion destruction path, continuum field theories can be used to obtain quantitative (and even analytical) values for the energy barriers. 
Here we argue that forth-order derivative terms, which are usually not considered when using, e.g., a micromagnetic model, determine that nature of the saddle point and thus the height of the energy barrier in an essential way. 
A continuum field theory or, equivalently, a variant of micromagnetic simulations, which includes effects of 4th-order derivatives, can be used to calculate the energy barriers provided that the spin-stiffness is positive and that 4th-order terms have a negative sign.
We will investigate two rather different types of skyrmions which are destroyed via two completely different singular spin configurations: 
skyrmions stabilized by Dzyaloshinskii-Moriya interactions (DMI) and skyrmions in frustrated magnets.

Due to the importance of skyrmion stability, many studies have investigated the creation and annihiliation of skyrmions and the height of the energy barrier \cite{bessarab2015method,hagemeister2015stability,rybakov2015newtype,
rohart2016path,lobanov2016mechanism,siemens2016minimal,yin2016topological,stosic2017paths,
cortesortuno2017thermal,bessarab2017lifetime,malottki2017enhanced,
uzdin2018energy,uzdin2018effect,leonov2018homogeneous,varentsova2018interplay,
desplat2018thermal,muller2018duplication,haldar2018first,derraschouk2018quantum,
garanin2018writing,buttner2018theory,wang2018atheory,bernandmantel2018theskyrmionbubble,
desplat2019paths,derraschouk2019thermal,malottki2019skyrmion,meyer2019isolated,potkina2019antiskyrmions,vlasov2019calculations}. 
From a methods perspective, perhaps the most important numerical approach to calculate energy barriers in classical spin systems is the geodesic nudged elastic band (GNEB) method pioneered by Bessarab \textit{et al.} \cite{bessarab2015method}, where one energetically optimizes spin configurations which smoothly connect an initial state (e.g. a magnetic skyrmion) and a final state (e.g. the ferromagnet). 
The highest-energy intermediate state, the saddle point, determines the energy barrier. 
Most results are availiable for skyrmions stabilized by DM interaction. 
Here it was found that, typically, skyrmions decay via shrinking to the size of a lattice constant\cite{rohart2016path} if they cannot escape via the edge of the system\cite{cortesortuno2017thermal,bessarab2017lifetime} or decay via quantum mechanical tunnelling\cite{vlasov2019calculations}.
Furthermore, two distinct concepts were identified that lead to higher energy barriers and, hence, more stable skyrmions: 
larger energy barriers were obtained for larger skyrmions\cite{varentsova2018interplay,potkina2019antiskyrmions}
while the energy barrier of smaller skyrmions was found to be increased by frustrating %(c.f., \textit{competing})
 next-nearest neighbor exchange interactions\cite{malottki2017enhanced}.
For a particular case of the latter scenario, in contrast to the symmetric shrinking path, an asymmetric decay path was reported\cite{meyer2019isolated} where the skyrmion annihilates via a singular defect at an off-center position.
Similar paths with off-centered and on-centered defects were reported by Desplat \textit{et al.}\cite{desplat2019paths} for a model system where skyrmions were stabilized by frustrated interactions on a square lattice without DMI.
Moreover, other methods have been employed to calculate the energy barrier for the decay of a skyrmion, including stochastic simulations\cite{schuette2014inertia,hagemeister2015stability,siemens2016minimal,rohart2016path,yin2016topological,derraschouk2019thermal} or a forced decay by increasing the external field\cite{derraschouk2018quantum,derraschouk2019thermal}.
Also attempts have been made to predict the energy barrier by approximate analytical solutions of the micromagnetic model which can be used to gather information about the isolated skyrmion\cite{buttner2018theory,bernandmantel2018theskyrmionbubble}.
However, as we show in this paper, the fourth order gradient corrections beyond the standard micromagnetic model  are essential to not only describe the energy barrier but also determine whether the physics of the barrier are universal or depend on the complicated microscopic details. 
A recent study by Derras-Chouk, Chudnovsky, and Garanin \cite{derraschouk2019thermal}, which was performed in parallel to our investigations, recognizes the importance of 4th-order terms for the calculation of thermal activation barriers.
We will discuss similarities and differences to the analysis of Ref.~\citenum{derraschouk2019thermal} below.

Experimentally, the most direct way to measure the energy barrier $\Delta E$ is to investigate the temperature dependence of the lifetime of skyrmions\cite{wild2017entropy,wilson2019measuring}, which is expected to be proportional to $e^{-\Delta E/T}$. 
For skyrmions in rather thick films of Cu$_2$OSe$_3$O it was shown\cite{wild2017entropy} that the activation barriers are determined by the emergence of a Bloch-point, i.e., a three-dimensional topological defect which caps the end of a skyrmion string.
Furthermore, the pre-exponential factor of the N\'eel-Arrhenius law was shown to induce major corrections of $40$ orders of magnitude to the lifetime.
For thin layers, however, this huge effect is not confirmed experimentally.

In the following, we will first consider a microscopic model for skyrmions stabilized by Dzyaloshinskii-Moriya interactions and investigate under which condition  and in which sense its collapse can be described by a continuum field theory. 
A second chapter will focus on skyrmions in frustrated magnets where skyrmion annihilation is governed by a completely different process.

\section{Skyrmions stabilized by Dzyaloshinskii-Moriya interactions}
\label{sec:dmi}
%----------------------------------------------------------------------------------------

%----------------------------------------------------------------------------------------
%----------------------------------------------------------------------------------------
\subsection{A minimal atomistic model}
\label{sec:dmi:model}
%----------------------------------------------------------------------------------------

\begin{figure}
\centering{
    \includegraphics[width=0.47 \textwidth]{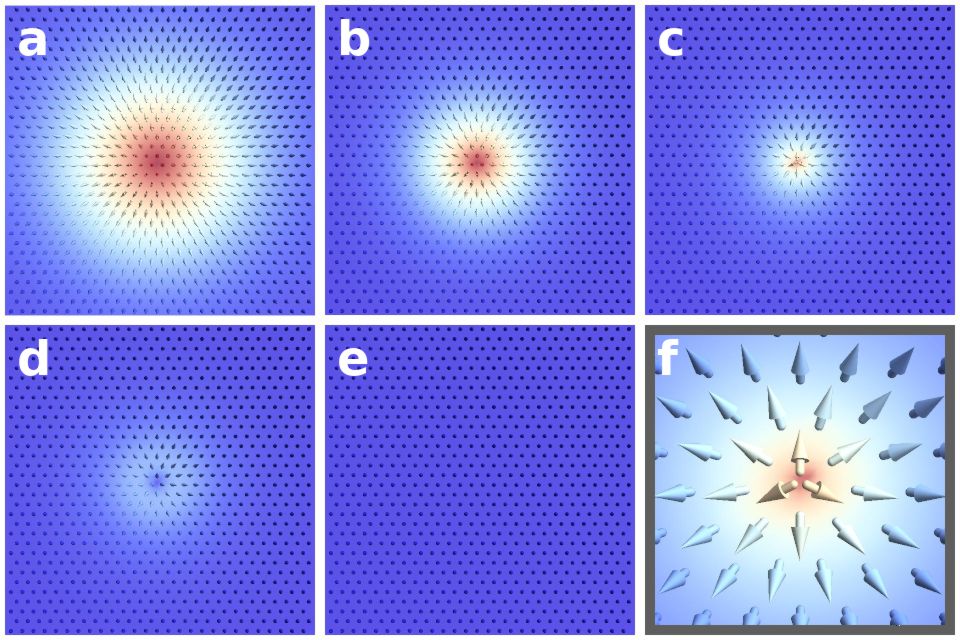}
    \vspace{1mm} 

    \includegraphics[width=0.47 \textwidth]{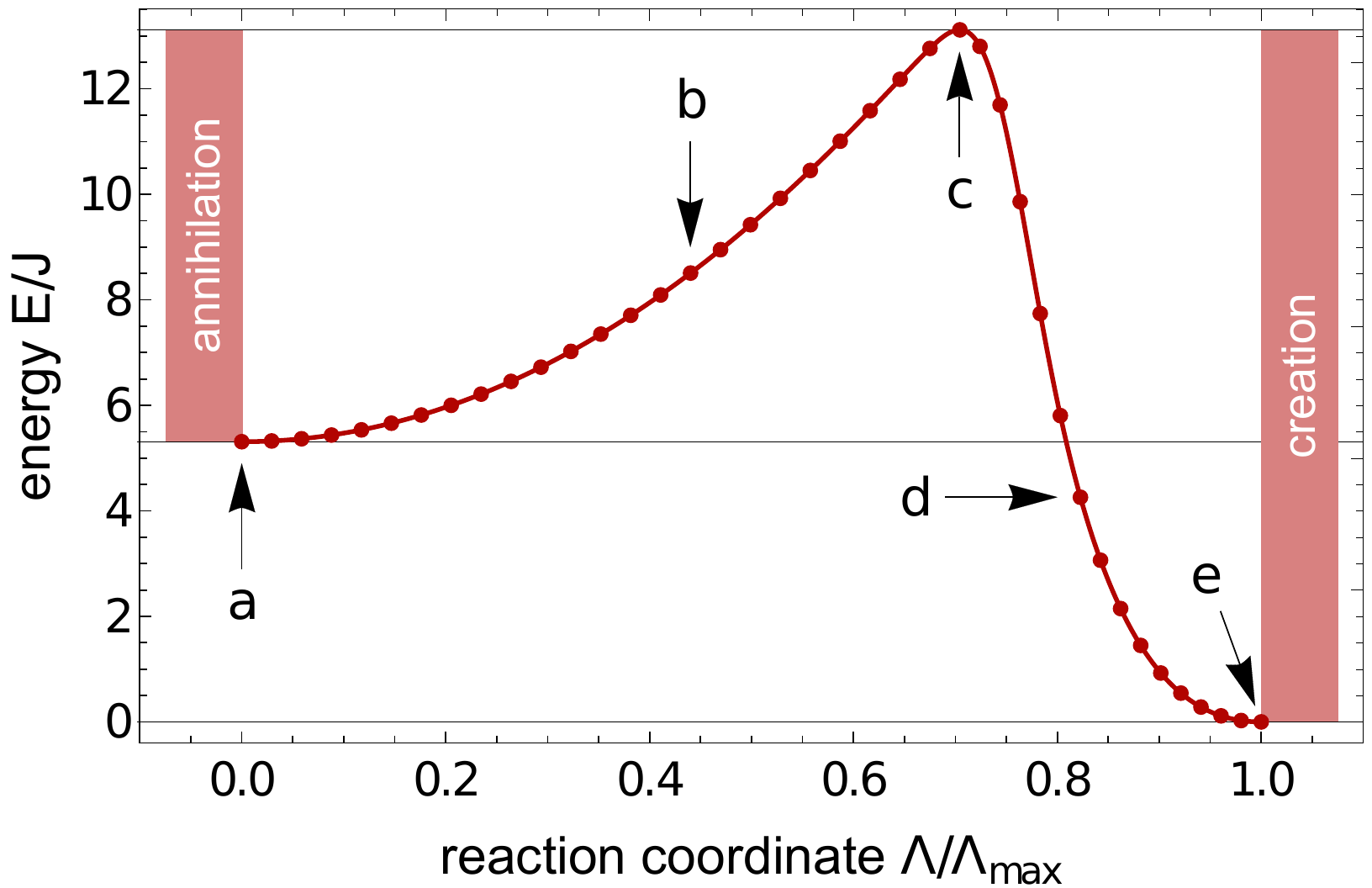}
    \caption{
        Minimal energy path for the creation/annihilation of a DMI-stabilized skyrmion as obtained from the GNEB method, see Sec.~\ref{sec:appendix:minimalenergypath}.
        The upper panels (a-f) show the real-space magnetic texture in the proximity of the center of the skyrmion for various states along the minimal energy path.
        The color encodes the out-of-plane component of the magnetization.  
        Panel (f) is a close-up of the saddle point texture in panel (c).
        The lower panel shows the energy $E/J$ along the minimal energy path as a function of the reaction coordinate $\Lambda$, see Sec.~\ref{sec:appendix:minimalenergypath}.
        The energy is evaluated with respect to the polarized phase (e).
        The arrows indicate the location of the upper panels in the minimal energy path.
        % DATA HERE WAS FOR LANGEVIN (J,D,B,a)=(1,1,1,0.2)
        The results were obtained for $D/J\!=\!0.2$, $\mu_s H/\!J\!=\!0.06$, $K\!=\!0$, and $a=1$. 
    }
    \label{fig1}
}
\end{figure}

In the following, we will consider a minimal model with only nearest neighbor interactions for a system which stabilizes magnetic skyrmions by (interfacial) Dzyaloshinskii-Moriya interactions  in a system without inversion symmetry. Ferromagnetic nearest-neighbor spin-orbit interactions of strength $J$ favour a parallel alignment of spins, while the interfacial interfacial Dzyaloshinskii-Moriya interaction (DMI) parametrized by $D$ induces inhomogeneous magnetic textures. Furthermore, we consider the effects of an external magnetic field $H$, and a uni-axial anisotropy $K$.
The energy for magnetic moments $\vec{m}_i \!=\! \vec{M}_i/M$ located on lattice sites $\vec{r}_i$ then reads
\begin{equation}
\begin{split}
    E =
    & - J \sum_{\langle i, j \rangle} \vec{m}_i \!\cdot\! \vec{m}_j
    - D \sum_{\langle i, j \rangle} \hat{d}_{i\!j} \!\cdot\! \left( \vec{m}_i \!\times\! \vec{m}_j \right) \\
    & - \mu_s H \sum_i m_i^z
    - K \sum_i (m_i^z)^2 \quad.
\end{split}
\label{eq:dmi:model:discrete}
\end{equation}
Here, $\langle i, j \rangle$ are nearest neighbors and every bond is counted once.
The DMI-vector $\hat{d}_{i\!j} \!=\! \hat{z} \!\times\! \frac{\vec{r}_i\!-\!\vec{r}_j}{|\vec{r}_i\!-\!\vec{r}_j|}$ fixes the sense of rotation for spin spirals, resulting in N\'eel-type skyrmions, see Fig.~\ref{fig1}a.
We will consider both square and triangular lattices in the following. In both cases, the lattice constant is given by $a$. 

For calculations of the minimal energy path, we use the geodesic nudged elastic band method\cite{bessarab2015method} (GNEB), see Sec.~\ref{sec:appendix:minimalenergypath}. This method allows to construct a minimal energy path (MEP) linking the skyrmion state to a ferromagnetic state. 
The maximum of the MEP defines the activation barrier for skyrmion annihiliation or creation.
As an example, a MEP for a given parameter set on a triangular lattice is shown in Fig.~\ref{fig1}. The snapshots of the spin-configuration obtained along the path, Fig.~\ref{fig1}a-f, show that the decay can be described as the shrinking of the skyrmion along a so-called collapse path \cite{bessarab2015method,rohart2016path,lobanov2016mechanism,cortesortuno2017thermal,bessarab2017lifetime,varentsova2018interplay,desplat2018thermal}.
Our goal will be to obtain an analytic understanding of the value of the energy at the saddle point configuration,
 shown in Fig.~\ref{fig1}c and as a close-up in Fig.~\ref{fig1}f, respectively.

\begin{figure}
    \center
    \includegraphics[width=0.47 \textwidth]{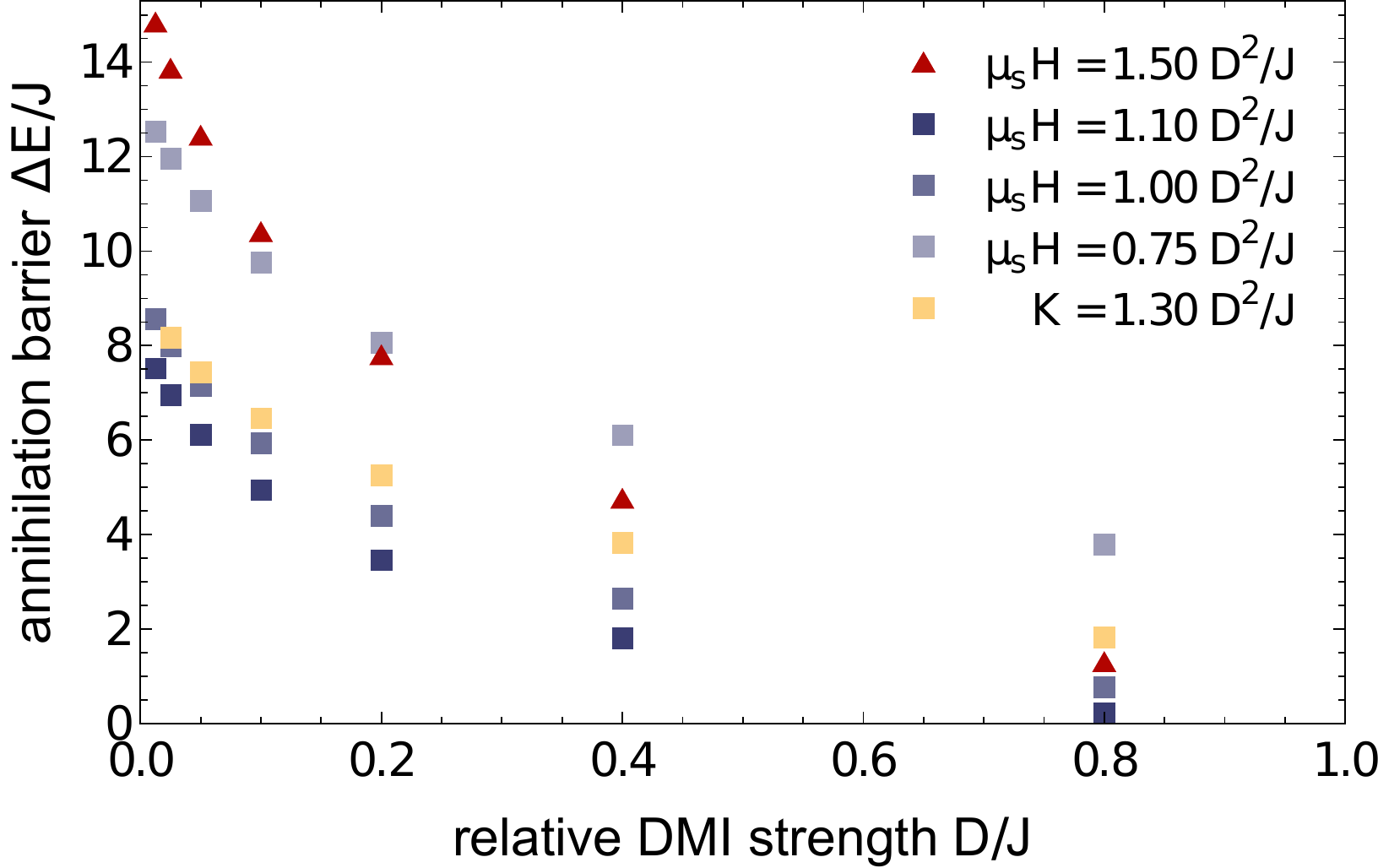}
    \caption{
        Energy barriers for the annihilation of an isolated skyrmion in the polarized state, c.f. Fig.~\ref{fig1}.
        The barrier is defined as the energy difference between the saddle point of the minimal energy path and the isolated skyrmion.
        The plot shows the results for a triangular lattice with a magnetic field $H$ (red) and a square lattice with a magnetic field $H$ (blue) or a uni-axial anisotropy $K$ (yellow), each for various values of Dzyaloshinskii-Moriya interaction $D$ and $a=1$.
        Data of the same color collapses onto the same lowest order effective theory for large skyrmions, see Eq.~\eqref{eq:dmi:model:continuous:rescaled}. 
    }
    \label{fig2}
\end{figure}

Our numerical studies confirm the collapse path mechanism also for other parameter sets on both the square and triangular lattice geometries. The resulting energy barriers for the annihilation of a skyrmion are summarized in Fig.~\ref{fig2}.
In agreement with the previous study by Varentsova \textit{et al.}\cite{varentsova2018interplay}, we observe that the annihilation barrier becomes larger when the Dzyaloshinskii-Moriya interaction $D/J$ is smaller.
As $J/D$ determines the radius of the skyrmion, this also implies that larger skyrmions are more stable.

In the following, we will derive an expression for the asymptotic behavior of the energy barrier.

%----------------------------------------------------------------------------------------
%----------------------------------------------------------------------------------------
\subsection{An effective continuum theory for large skyrmions}
\label{sec:dmi:largeskyrmions}
%----------------------------------------------------------------------------------------

In order to study the asymptotic limit for very large skyrmions, we derive an effective continuum theory for the atomistic energy functional, Eq.~\eqref{eq:dmi:model:discrete}.
Therefore, we write the magnetization on the discrete lattice with lattice constant $a$ in a Fourier representation
\begin{equation}
    \vec{m}(\vec{r}) = \frac{1}{\sqrt{V_q}} \int\!\mathrm{d}^2q \,\,\tilde{\vec{m}}(\vec{q}) e^{i \vec{q}\cdot\vec{r}}
\label{eq:dmi:model:fourier}
\end{equation}
and subsequently apply an approximation for smooth variations of the magnetization, $|\vec{q}|\ll a^{-1}$, where we, in a first step, consider all terms up to quadratic order $\mathcal{O}(q^2)$.
After applying the continuous back-transformation into real-space again, the leading-order gradient expansion of the energy functional reads
\begin{equation}
\begin{split}
    \delta E[\vec{m}] = \!\int\!\!\mathrm{d}^2r \,\,
    & \frac{\mathcal{J}}{2} (\nabla \vec{m})^2 - \mathcal{D} \,\vec{m} \cdot \left(\left(\hat{z}\!\times\!\nabla\right)\!\times\!\vec{m}\right) \\
    & - \mu_0 \mathcal{H} (m_z-1) - \mathcal{K} (m_z^2-1)
\end{split}
\label{eq:dmi:model:continuous}
\end{equation}
with the effective continuum interaction constants $\mathcal{J}\!=\!J$, $\mathcal{D}\!=\!D/a$, $\mathcal{H}\!=\!H/a^2$, and $\mathcal{K}\!=\!K/a^2$ for the square lattice, or  $\mathcal{J}\!=\!\sqrt{3}J$, $\mathcal{D}\!=\!\sqrt{3}D/a$, $\mathcal{H}\!=\!2 H/(\sqrt{3}a^2)$, and $\mathcal{K}\!=\!2 K/(\sqrt{3}a^2)$ for the triangular lattice, respectively.
Note that this second order effective theory is independent of the geometry of the lattice and obeys a continuous symmetry for rotations around the $\hat{z}$-axis.
In particular, the triangular lattice with $\mu_s H=1.50 D^2/J$ and the square lattice with $\mu_s H=1.00 D^2/J$ assume the same solution in this limit and all skyrmion solution are circular symmetric.

From a dimensional analysis, we find that the continuum model has the intrinsic length scale $\xi=\mathcal{J}/\mathcal{D}=a J/D$.
By rewriting the spatial distances in units of this length scale, $\vec{r} = \xi \tilde{\vec{r}}$, one obtains the dimensionless energy functional
\begin{equation}
\begin{split}
    \delta E/\mathcal{J} = \!\int\!\!\mathrm{d}^2\tilde{r} \,\,
    & \frac{1}{2} (\tilde{\nabla} \vec{n})^2 - \vec{n} \cdot \left(\left(\hat{z}\!\times\!\tilde{\nabla}\right)\!\times\!\vec{n}\right) \\
    & - h (n_z-1) - \kappa (n_z^2-1)
\end{split}
\label{eq:dmi:model:continuous:rescaled}
\end{equation}
with $h=\mu_0 \mathcal{J}\mathcal{H}/\mathcal{D}^2 $ and $\kappa= \mathcal{J}\mathcal{K}/\mathcal{D}^2$ where for convenience we use $\vec{n}(\tilde{\vec{r}})=\vec{m}(\vec{r})$.

The scaling analysis reveals two important features of skyrmions in the continuum limit:
(i) for the same set of effective parameters ($h$,$\kappa$) they are described by the same solution $\vec{n}(\tilde{\vec{r}})$ and
(ii) the size of the skyrmion is linear in $\xi=\mathcal{J}/\mathcal{D}$ for fixed $h$ and $\kappa$.
In the following, we will investigate corrections to the continuum limit and derive an expression for the saddle point of the collapse path for large skyrmions.

\subsubsection{Corrections to the continuum theory}
\label{sec:dmi:largeskyrmions:skyrmionenergy}

The above continuum theory, Eq.~\eqref{eq:dmi:model:continuous:rescaled}, was derived only up to quadaratic order. Corrections will, however, be important, especially when the skyrmion shrinks along the collapse path.
We can derive the leading order correction terms $\delta E^{(4)}$ to the quadratic continuum theory by repeating the derivation sketched above where we take into account all terms up to fourth order $\mathcal{O}(q^4)$ instead.

In contrast to the $\mathcal{O}(q^2)$-theory, Eq.~\eqref{eq:dmi:model:continuous}, the fourth order terms depend on the lattice geometry.
For the square lattice, the expression takes the form
\begin{equation}
    \delta E^{(4)}_\text{\tiny{$\square$}} = -\!\int\!\!\mathrm{d}^2r \,\,
    \frac{J a^2}{24} \left(\partial_\alpha^2 \vec{m}\right)^2     
    + \frac{D a}{6} \, \vec{m} \!\cdot\! \left(\!\left(\hat{z}\!\times\!\hat{e}_\alpha\partial^3_\alpha\right)\!\times\!\vec{m}\right)
\label{eq:dmi:model:continuous:correction:square}
\end{equation}
with implicit summation over the spatial index $\alpha=x,y$ and we use the unit vectors $\hat{e}_x=\hat{x}$ and $\hat{e}_y=\hat{y}$.
This expression inherits the discrete $90^\circ$ rotation symmetry of the square lattice.
The correction term for the triangular lattice geometry, in turn, reads 
\begin{equation}
\begin{split}
    \delta E^{(4)}_\text{\tiny{$\triangle$}} = &-\!\int\!\!\mathrm{d}^2r \,\,
    \frac{\sqrt{3}\, J a^2}{32} \left(\nabla^2 \vec{m}\right)^2 \\
    &+ \frac{\sqrt{3}\, D a}{ 8} \left(\nabla^2 \vec{m}\right) \!\cdot\! \left(\!\left(\hat{z}\!\times\!\nabla\right)\!\times\!\vec{m}\right)
\end{split}
\label{eq:dmi:model:continuous:correction:triangle}
\end{equation}
with the Laplace operator $\nabla^2=\partial^2_x+\partial^2_y$. Note that here the rotation symmetry is still continuous.

Using the rescaled coordinates of Eq.~\eqref{eq:dmi:model:continuous:rescaled}, the leading corrections to the continuum theory of a circluarly symmetric skyrmion can in general be written as
\begin{equation}
\begin{split}
    \frac{\delta E^{(4)}}{\mathcal J} = \!\int\!\!\mathrm{d}^2r\,\,
        &\frac{\mathcal{K}_4}{2} \!\left(\!\tilde{\nabla}^2 \vec{n}\!\right)^2
        \!\!+ \mathcal{K}_3  \left(\!\tilde{\nabla}^2\vec{n}\!\right) \!\cdot\! \left(\!\!\left(\!\hat{z}\!\times\!\tilde\nabla\!\right)\!\times\!\vec{n}\!\right) \\
    + &\frac{\mathcal{K}_4'}{4} \left(\!\left(\tilde{\nabla} \vec n\right)^2\right)^2 
        \!\!+ \frac{\mathcal{K}_4''}{4} \left(\!\tilde{\partial}_\alpha \vec n \cdot \tilde{\partial}_{\alpha'} \vec n\!\right)^2
\end{split}
\label{eq:dmi:model:continuous:correction}
\end{equation}
where the (negative) values of  $\mathcal{K}_4=-\frac{\mathcal{D}^2 a^2}{16 \mathcal{J}^2}$ and $\mathcal{K}_3=-\frac{\mathcal{D}^2 a^2}{8 \mathcal{J}^2}$ are accidentially the same for the square and the triangular lattice when expressed in terms of the coupling constants of the continuum theory. 
The prefactor of the quartic 4-spin terms vanish for our microscopic models, $\mathcal{K}_4'=\mathcal{K}_4''=0$. 
We have included them neverthess here as they may exist in real materials, especially in metallic compounds \cite{heinze2011spontaneous}. 
As they have the same scaling properties as the  $\mathcal{K}_4$ term, they turn out to be equally important for the calculation of the activation barriers.
For the square lattice, the rotationally invariant term $(\!\tilde{\nabla}^2 \vec{n}\!)^2$ has been obtained from $(\partial_\alpha^2 \vec{m})^2 $ by averaging over spatial directions. 
A corresponding correction term proportional to $(\partial_\alpha^2 \vec{n})^2  -\frac{3}{2}(\!\tilde{\nabla}^2 \vec{n})^2$ has been omitted in Eq.~\eqref{eq:dmi:model:continuous:correction} as its expecation value vanishes for rotationally invariant textures. 
Therefore it does not contribute to linear order perturbation theory and gives only subleading corrections.

\begin{figure}
    \center
    \includegraphics[width=0.47 \textwidth]{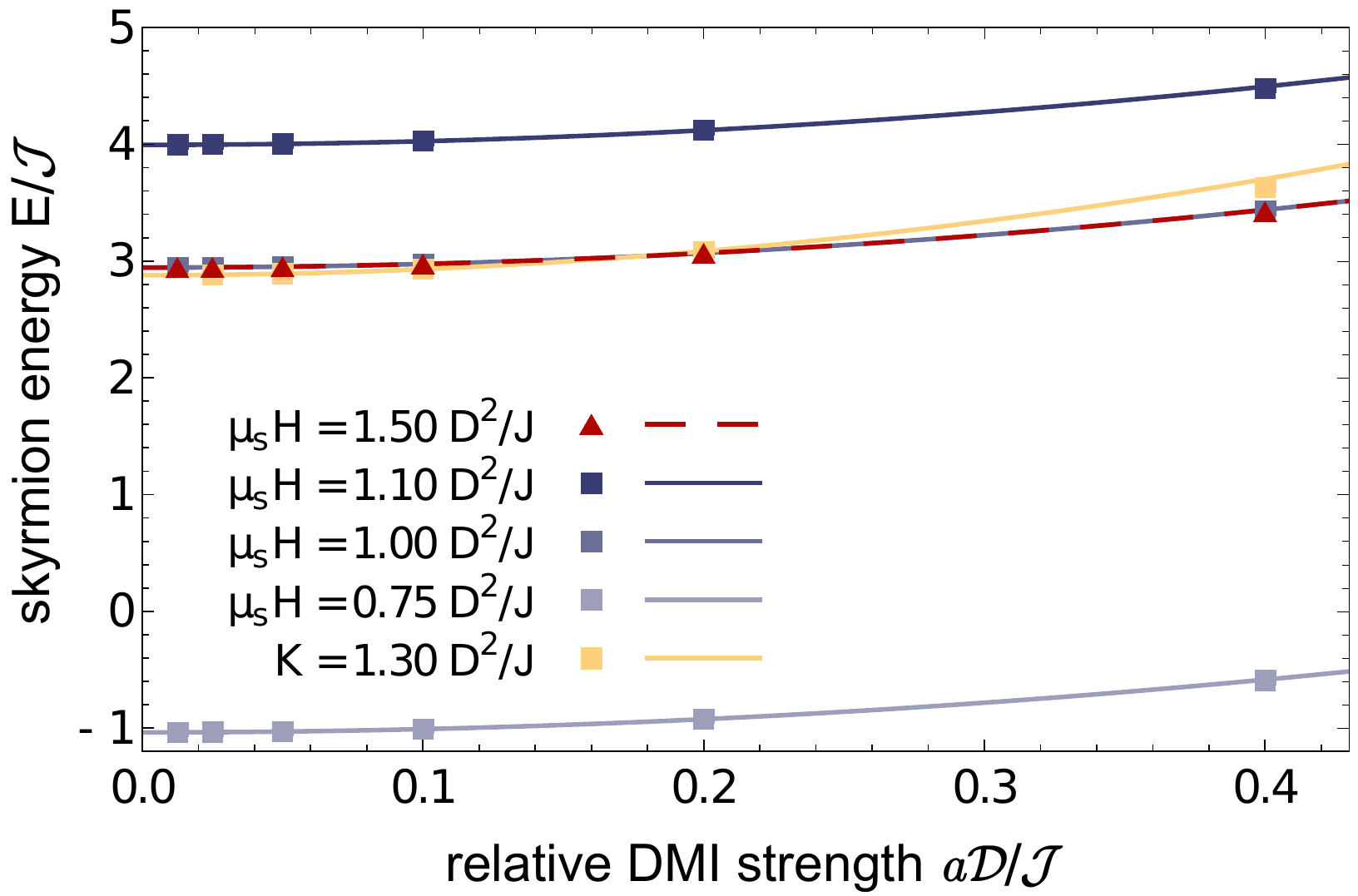}
    \caption{
        Energy of an isolated skyrmion relative to the polarized state.
        The plot shows the results obtained for the lattice model Eq.~\eqref{eq:dmi:model:discrete}, evaluated on a triangular lattice with a magnetic field $H$ (red triangles) and a square lattice with a magnetic field $H$ (blue squares) or a uni-axial anisotropy $K$ (yellow squares), each for various values of Dzyaloshinskii-Moriya interaction $D$.
        The results are plotted in the units of the effective continuum theory, Eq.~\eqref{eq:dmi:model:continuous}.
        The parabolic lines are the results of this effective theory with the leading order perturbative corrections, see Eq.~\eqref{eq:dmi:model:continuous:correction}.
        The dashed red line and the solid blue line with $\mu_s H=1.00 D^2/J$ are identical within this approximation.
    }
    \label{fig3}
\end{figure}

Using that $|\mathcal{K}_{3/4}|\ll 1$ for skyrmions, which are much larger than the lattice constant, $\mathcal{J}/\mathcal{D}\gg a$, we can calculate skyrmion energy up to linear order in $\mathcal{K}_{3/4}$, i.e., up to order $(a/\xi)^2$ by first computing the rotationally-symmetric minimum of the leading-order energy functional Eq.~\eqref{eq:dmi:model:continuous:rescaled}, $\vec{n}_s(\tilde{\vec{r}})=\hat{r} \sin\theta_s(\tilde{r}) + \hat{z} \cos\theta_s(\tilde{r})$. The leading-order correction due the resulting energy is then computed by evaluating $\delta E^{(4)}[\vec{n}_s(\tilde{\vec{r}})]$, Eq.~\eqref{eq:dmi:model:continuous:correction}, for this skyrmion texture.
A comparison of this perturbative result and the result of the lattice model, Eq.~\eqref{eq:dmi:model:discrete}, is shown in Fig.~\ref{fig3}. 
As expected, the two distinct approaches are in very good agreement and the corrections to the leading order continuum description are suppressed by the tiny factor $(a/\xi)^2\sim (\mathcal{J}/\mathcal{D})^2$ for skyrmions which are much  larger than the lattice spacing. 
As in most experimental systems the relativistic spin-orbit interaction is much weaker than the magnetic exchange interaction, hence, this is also the experimentally relevant limit.

We can, however, expect that the corrections are much more important for the computation of the energy barrier along the collapse path of the skyrmion.
    
\subsubsection{The saddle point of the collapse path}
\label{sec:dmi:largeskyrmions:saddlepoint}

In Fig.~\ref{fig1} we have shown that the skyrmion shrinks before it collapses as has been discussed in numerous previous studies \cite{bessarab2015method,rohart2016path,
lobanov2016mechanism,cortesortuno2017thermal,bessarab2017lifetime,varentsova2018interplay,desplat2018thermal}.
The main reason for the collapse is that the destruction of a skyrmion necessarily involves a singular, vortex-like spin configuration. 
As discussed in more detail in Sec.~\ref{sec:frustration}, a vortex-like solution of radius $R_v \gg a$ is associated with an energy cost approximately given by $\pi \mathcal{J} \ln[R_v/a]$. 
This energy cost can be minimized by making $R_v$ smaller and smaller. 
Along the collapse path, the skyrmion therefore shrinks in size and the vortex is formed for $R_v \sim a$.  
We will show below that the saddle point, i.e. the energetic bottleneck along the collapse path, is for a wide class of systems obtained from a very small skyrmion and not from the vortex solution.
Moreover, for sufficiently large initial skyrmions this small skyrmion is still much larger than the underlying atomic lattice and, thus, can be described by a continuum theory.

We first analyze skyrmions with radius $R_s$ for $a \ll R_s \ll \xi$. 
This is done most conveniently by introducing dimensionless coordinates $\vec x$ with $\vec r= R_s \vec x$ or, equivalently, $\tilde{\vec r}=\frac{R_s}{\xi} \vec x$.
In these units we find
\begin{equation}
 \begin{split}
  &\frac{\delta E}{\mathcal{J}} \approx \!\int\!\!\mathrm{d}^2 x\,\,
     \frac{1}{2} \left({\nabla}\vec{n}\right)^2 
     - \frac{R_s}{\xi} \, \vec{n} \!\cdot\! \left(\hat{z}\!\times\!{\nabla}\right)\!\times\!\vec{n} \\
  &+ \left(\!\frac{\xi}{R_s}\!\right)^2 \!\! \left(\! \frac{\mathcal{K}_4}{2}  \!\left({\nabla}^2 \vec{n}\right)^2 + \!\frac{\mathcal{K}_4'}{4} (({\nabla} \vec n)^2)^2 + \!\frac{\mathcal{K}_4''}{4} ({\partial}_\alpha \vec n  \!\cdot\! {\partial}_{\alpha'} \vec n)^2\!\right) \\
 & +\mathcal{O}\!\left(\left(\!\frac{R_s}{\xi}\!\right)\!^2, \,\mathcal{K}_3 \frac{\xi}{R_s} \right).
 \end{split}
\label{eq:BP:correction}
\end{equation}
The DM interaction is suppressed by the small prefactor $R_s/\xi$ and also the higher-order gradient terms are small
as $\mathcal{K}_4 (\xi/R_s)^2 \sim (a/R_s)^2 \ll 1$ for $R_s \gg a$. 
Therefore, the main contribution to the energy arises for $a \ll R_s \ll \xi$ from the scale-invariant term
\begin{equation}
    \frac{\delta E}{\mathcal{J}} \approx \int\!\!\mathrm{d}^2\tilde{\vec r}\,
     \frac{1}{2} ({\nabla}\vec{n})^2, \label{eq:BP}
\end{equation}
where we now use again the dimensionless variables $\tilde r$ of Eq.~\eqref{eq:dmi:model:continuous:rescaled} and \eqref{eq:dmi:model:continuous:correction}.

A solution which minimizes the energy~\eqref{eq:BP} for fixed skyrmion winding number was found by Belavin and Polyakov~\cite{belavin1975metastable}. 
This Belavin-Polyakov skyrmion $\vec{n}_{\text{\tiny{BP}}}(\tilde{\vec{r}})$ can be expressed as
\begin{equation}
    \vec{n}_{\text{\tiny{BP}}}(\tilde{\vec{r}}) = \hat{r} \sin \theta_0(\tilde{r}) + \hat{z} \cos \theta_0(\tilde{r}), \,\,\, \theta_0(\tilde{r})=2\arctan\!\left(\!\frac{\tilde{r}_{\text{\tiny{BP}}}}{\tilde{r}}\!\right)
\label{eq:dmi:continuous:ultrasmall:polyakovskyrmion}
\end{equation}
where $\tilde{r}_{\text{\tiny{BP}}} = r_{\text{\tiny{BP}}}/\xi$ is the radius of this skyrmion, i.e., where the magnetization is in-plane.
The resulting energy of the Belavin-Polyakov skyrmion with respect to the polarized state evaluates to
\begin{equation}
    \delta E_0[\vec{n}_{\text{\tiny{BP}}}]/\mathcal{J} = 4 \pi 
\label{eq:dmi:continuous:ultrasmall:energy4pi}
\end{equation}
and is independent of the skyrmion radius $ r_{\text{\tiny{BP}}}$. 
We therefore have to compute the leading order correction by evaluating all terms in Eq.~\eqref{eq:BP:correction} 
using the Belavin-Polyakov solution.

The evaluation of the correction from the DM interaction turns out to be non-trivial. 
A straightforward integration gives
 $\int\! d^2 \tilde r\,\, \vec{n}_{\text{\tiny{BP}}} \!\cdot\! (\hat{z}\!\times\!\tilde{\nabla})\!\times\!\vec{n}_{\text{\tiny{BP}}}= 4 \pi r_{\text{\tiny{BP}}}$. 
 While this integral is well converging, one has to take into account a mathematical subtlety (apparently overlooked in Ref.~[\onlinecite{derraschouk2019thermal}]) which arises because the Belavin-Polyakov decays very slowly (proportional to $1/\tilde r$) for large distances towards the ferromagnetic state. 
In the presence of  anisotropy terms or external magnetic fields, the real skyrmion decays, however, faster in this limit. 
This effect can be described by multiplying the $\theta_0(\tilde r)$ in Eq.~\eqref{eq:dmi:continuous:ultrasmall:polyakovskyrmion} with a smooth cutoff function $f_c(\tilde r)$ with $f_c(\tilde r \ll \tilde r_c)=1$ and $f_c(\tilde r \gg \tilde r_c)=0$, where $\tilde r_c\gg r_{\text{\tiny{BP}}}$ is a cutoff length much larger
than the radius of the  Belavin-Polyakov  skyrmion. With this cutoff, the integral obtains an extra contribution $-4 \pi \tilde r_{\text{\tiny{BP}}} \int_0^\infty dx\, f_c'(x)=4 \pi r_s (f_c(0)-f_c(\infty))=4 \pi r_{\text{\tiny{BP}}}$ in this limit. While this contribution vanishes for the idealized Belavin-Polyakov skyrmion ($f_c=1$), it changes the value of the integral from $4 \pi r_{\text{\tiny{BP}}}$ to $8 \pi r_{\text{\tiny{BP}}}$ for any finite cutoff function.

\begin{figure}
    \center  
    \includegraphics[width=0.47 \textwidth]{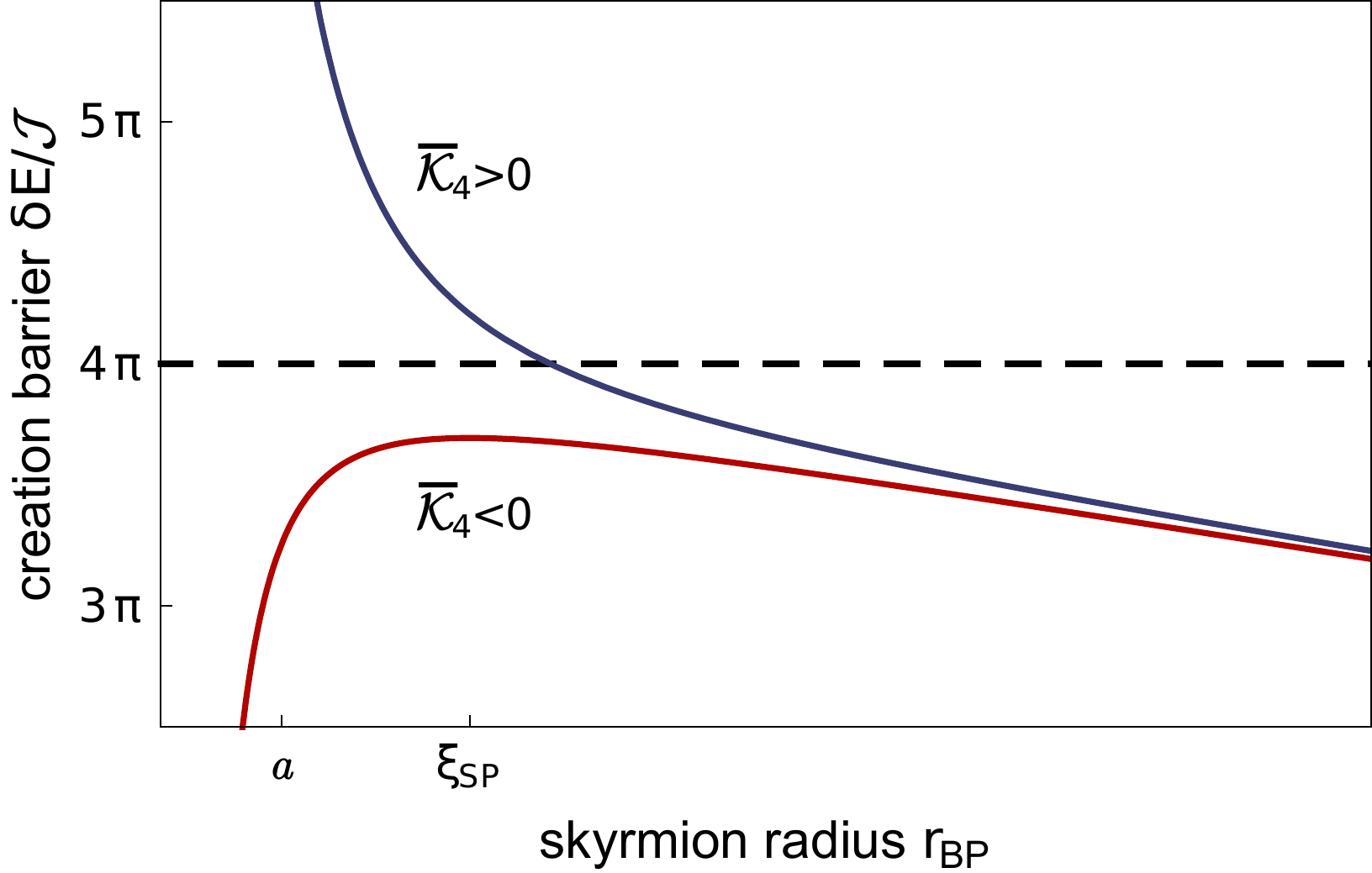}
    \caption{Schematic plot of the skyrmion energy as function of the skyrmion radius based on Eq.~\eqref{eq:dmi:continuous:ultrasmall:perturbation:energyfunctional:approx}. For $\bar{\mathcal{K}}_4<0$ the maximum with an energy {\em smaller} than $4 \pi \mathcal J$ occurs for $r_{\text{\tiny{BP}}}>a$ and is described by the continuum theory in Eq.~\eqref{eq:BP:correction}. For $\bar{\mathcal{K}}_4>0$, in contrast, the barrier is typically larger than $4 \pi \mathcal J$ and is determined by physics at the scale of the lattice constant $a$ where the continuum theory breaks down.
    }
    \label{fig4} 
\end{figure}

Taking this effect into account, we obtain for $a \ll r_{\text{\tiny{BP}}} \ll \xi$ 
\begin{align}\label{eq:dmi:continuous:ultrasmall:perturbation:energyfunctional:approx}  
    \frac{\delta E}{\mathcal{J}} &\approx 
    4 \pi +\bar{\mathcal{K}}_4  \frac{32 \pi}{3} \frac{1}{\tilde{r}_{\text{\tiny{BP}}}^2} - 8 \pi \, \tilde{r}_{\text{\tiny{BP}}} \\
    \bar{\mathcal{K}}_4 &= \mathcal{K}_4+\frac{\mathcal{K}_4'}{2}+\frac{ \mathcal{K}_4''}{4}. \nonumber
\end{align}
A similar formula (up to prefactors as discussed above) 
was obtained very recently in Ref.~[\onlinecite{derraschouk2019thermal}].
The property of this function depends on the sign of  $ \bar{\mathcal{K}}_4$ as is shown in Fig.~\ref{fig4}.
We will first discuss the case of {\em negative} $ \bar{\mathcal{K}}_4 $ realized in the microscopic model of 
Eq.~\eqref{eq:dmi:model:discrete}.
For negative $  \bar{\mathcal{K}}_4$, the energy $\delta E$ has a maximum as a function of $\tilde{r}_{\text{\tiny{BP}}}$ which defines the saddle point of the collapse path, see Fig.~\ref{fig4}. We denote the size of the Belavin-Polyakov skyrmion at the saddle point by $\xi_{\text{\tiny{SP}}}$ and its energy relative to the ferromagnetic state by $\delta E_{\text{\tiny{SP}}}$ and obtain for $  \bar{\mathcal{K}}_4<0$
\begin{eqnarray}
\xi_{\text{\tiny{SP}}}&\approx & 2 \xi \left(-\frac{\bar{\mathcal{K}}_4}{3}\right)^{1/3}  \!=a\left(\frac{J}{6 D}\right)^{1/3}= a\left(\frac{\xi}{6 a}\right)^{1/3} \nonumber  \\
    \frac{\delta E_{\text{\tiny{SP}}}}{\mathcal{J}} &\approx& 
    4 \pi - 3^{2/3} 8 \pi \, (-\bar{\mathcal{K}}_4)^{1/3}= 4 \pi -2 \pi \left(6 \frac{ D}{J}\right)^{2/3}\nonumber \\&=& 4 \pi -2 \pi  \left(6 \frac{a}{\xi}\right)^{2/3} \label{eq:dmi:continuous:ultrasmall:perturbation:energy}
\end{eqnarray}
where we inserted the value of $\bar{\mathcal{K}}_4$ for the microscopic model, Eq.~\eqref{eq:dmi:model:discrete}, and used $\xi=\mathcal{J}/\mathcal{D}=a J/D$ as above. 
Note that for large skyrmions with  $D \ll  J/6 $, we find that $\xi_{\text{\tiny{SP}}} \gg a$ which formally justifies the use of the continuum model for the calculation of the saddle point. 
For larger skyrmions, the height of the barrier approaches the universal limit $4 \pi \mathcal{J}$.
The importance of this energy scale (without a quantitative discussion of the range of validity and of corrections) has  also been realized in a number of previous studies\cite{kiselev2011chiral,leonov2016properties,buttner2018theory,bernandmantel2018theskyrmionbubble}.

\begin{figure}
    \center  
    \includegraphics[width=0.47 \textwidth]{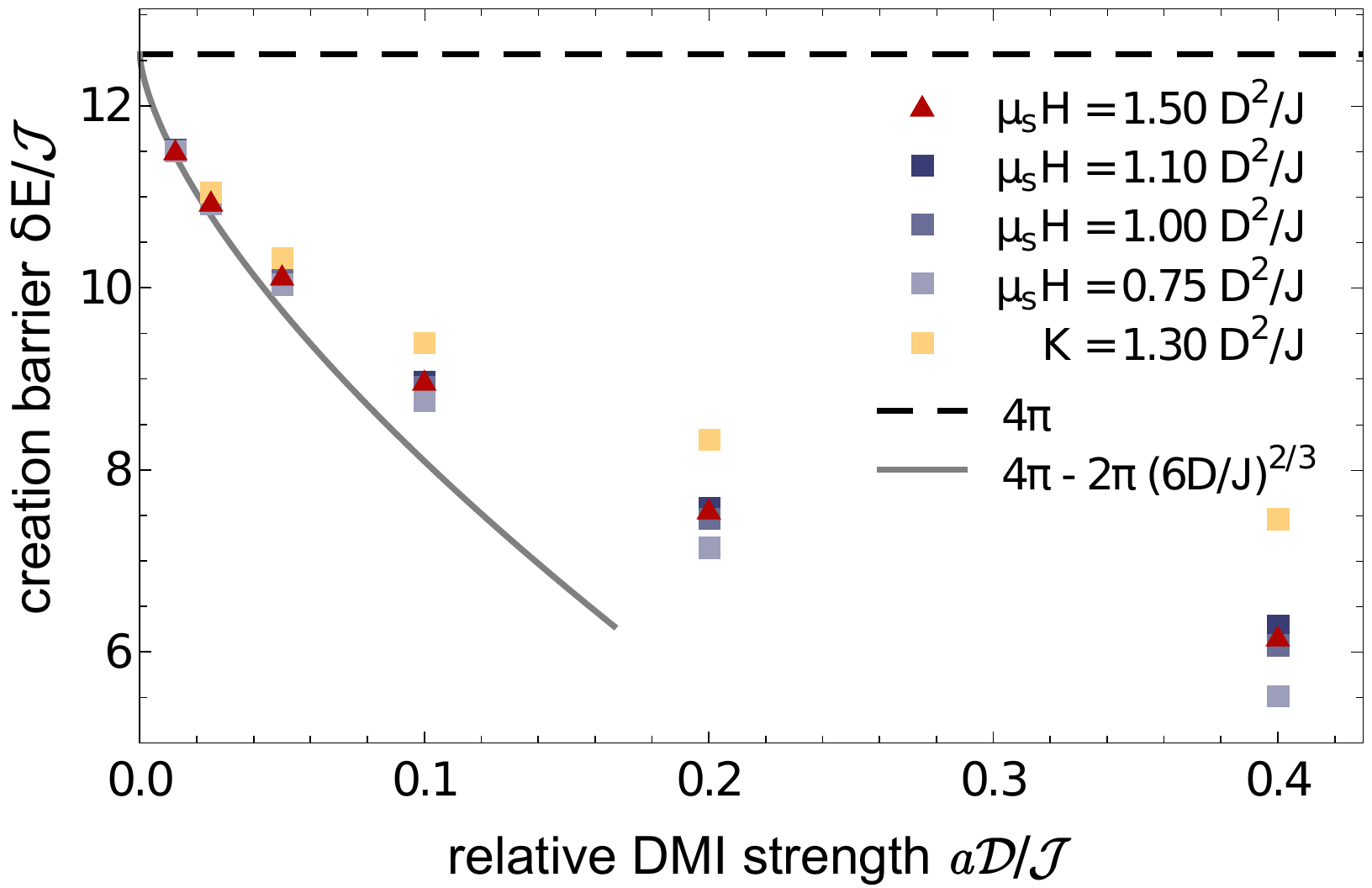}
    \caption{
        Creation barrier, i.e., energy of the saddle point, for the creation of a skyrmion in the polarized state, c.f. Fig.~\ref{fig1}.
        The plot shows the results obtained for the lattice model Eq.~\eqref{eq:dmi:model:discrete}, evaluated on a triangular lattice with a magnetic field $H$ (red triangles) and a square lattice with a magnetic field $H$ (blue squares) or a uni-axial anisotropy $K$ (yellow squares), each for various values of Dzyaloshinskii-Moriya interaction $D$.
        The results are plotted as a function of $a\mathcal{D}/\mathcal{J}=D/J=a/\xi$, see Eq.~\eqref{eq:dmi:model:continuous}, where $\xi$ is the skymrion size.
        The dashed black line indicates the analytical upper limit $\Delta E/\mathcal{J} = 4\pi$, see Sec.~\ref{sec:dmi:largeskyrmions:saddlepoint}, approached for $\xi \to \infty$. 
        The gray line is the perturbative analytical result, see Eq.~\eqref{eq:dmi:continuous:ultrasmall:perturbation:energy}, which is valid for  $a \mathcal{D}/\mathcal{J}\ll 1/6$.
    }
    \label{fig5} 
\end{figure}

In Fig.~\ref{fig5} we compare the analytical formula, Eq.~\eqref{eq:dmi:continuous:ultrasmall:perturbation:energy}, (grey line) to numerical results obtained from the GNEB method
 in the original atomistic model, Eq.~\eqref{eq:dmi:model:discrete}, for various values of anisotropy and magnetic field and for both the square and the triangular lattice. For $D \ll  J/6 $, a very good agreement is obtained, fully confirming the validity of the analytical formula for $\bar{\mathcal{K}}_4<0$. The comparison to the numerics clearly show that the energy bottleneck arises in a regime where the continuum theory is valid.

Note that the energy of the saddle point depends for large skyrmions only very weakly on the magnetic field and the anisotropy as they provide only subleading corrections to the saddle point energy. In contrast, the energy of the skyrmion does depend strongly on those parameters, see Fig.~\ref{fig3}, but is almost independent of $\bar{\mathcal{K}}_4$. As the annihilation barrier is given by the difference of saddle-point and skyrmion energy, the field dependence of the barrier height is dominated by the field dependence of the skyrmion energy.

\begin{figure}
    \center  
    \includegraphics[width=0.47 \textwidth]{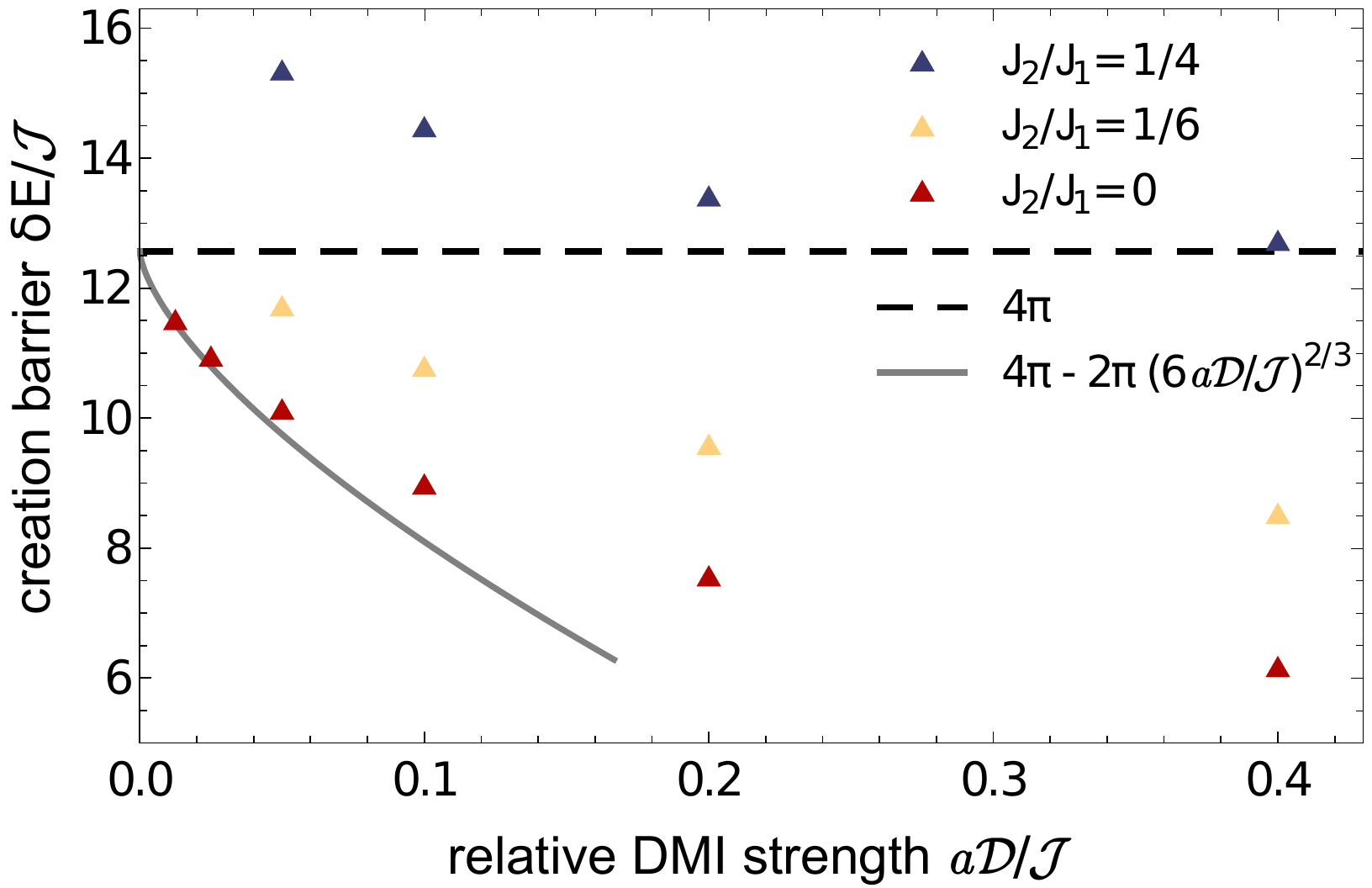}
    \caption{
        Creation barrier, i.e., energy of the saddle point, for the creation of a skyrmion in the polarized state, c.f. Fig.~\ref{fig1}, in the presence of frustrating interactions.
    The plot shows the results obtained for the lattice model Eq.~\eqref{eq:frustration:model}, evaluated on
   a triangular lattice for  $h=1$ and $\kappa=0$ in the dimensionless units of Eq.~\eqref{eq:dmi:model:continuous:rescaled} for three different values of $J_2/J_1$.
      While $\tilde{\mathcal{K}}_4=-\frac{a^2 \mathcal{D}^2}{16 \mathcal{J}^2}<0$ for $J_2=0$ (red triangles, same data as in Fig.~\ref{fig5}), we obtain
    $\tilde{\mathcal{K}}_4=\frac{a^2\mathcal{D}^2}{16 \mathcal{J}^2}>0$ for $J_2/J_1=1/6$ (yellow triangles) and 
 $ \tilde{\mathcal{K}}_4=5\frac{a^2 \mathcal{D}^2}{16 \mathcal{J}^2}>0$  for $J_2/J_1=1/4$ (blue triangles). In the latter case, the creation barrier is substantially larger than $4 \pi \mathcal J$.
}
    \label{fig6} 
\end{figure}

For a positive sign of $\bar{\mathcal{K}}_4$, the energy of the skyrmion as function of its radius  does not have a maximum for $r_{\text{\tiny{BP}}}>a$, see Fig.~\ref{fig4}. 
In this case, the energy of the saddle point can become larger than $4 \pi \mathcal{J}$ and determined by the microscopic physics at the length scale of the lattice constant $a$. 
In a metal, for example, one can expect that details of the electronic states scattering from a highly non-colinear local spin configuration will determine the energy barrier. 
Within our microscopic model, positive values of $\bar{\mathcal{K}}_4$ can be obtained by adding a frustrated antiferromagnetic next-nearest neighbor interaction to the microscopic model (details are discussed in the next section). In Fig.~\ref{fig6} we show the resulting energy barriers calculated with the GNEB method for frustrating next-nearest neighbor interactions which drive $\tilde{\mathcal{K}}_4$ positive. For sufficient strong frustration, the energy barrier becomes larger than $4 \pi \mathcal J$, see  Fig.~\ref{fig6}.

\section{Skyrmions in frustrated Heisenberg ferromagnets}
\label{sec:frustration}
%----------------------------------------------------------------------------------------

Recent theoretical studies proposed competing interactions as an alternative mechanism to stabilize magnetic skyrmions in symmetric magnetic layers\cite{ivanov1990magnetic,okubo2012multiple,leonov2015multiply,lin2016ginzburg,hayami2016bubble}.
Under certain conditions, skyrmions can exist in these systems as stable quasi-particles without the need for broken lattice symmetry and spin-orbit coupling.
The lack of broken inversion symmetry results in equal energies for N\'eel and Bloch skyrmions and all intermediate helicities and, furthermore, also for anti-skyrmions.

In the following, we will determine the minimal energy transition path from a skyrmion to the polarized state or an anti-skyrmion and develop an approximate expression for large skyrmions.

%----------------------------------------------------------------------------------------
%----------------------------------------------------------------------------------------
\subsection{A minimal atomistic model with inversion symmetry}
\label{sec:frustration:model}
%----------------------------------------------------------------------------------------

\begin{figure}
    \center
    \includegraphics[width=0.47 \textwidth]{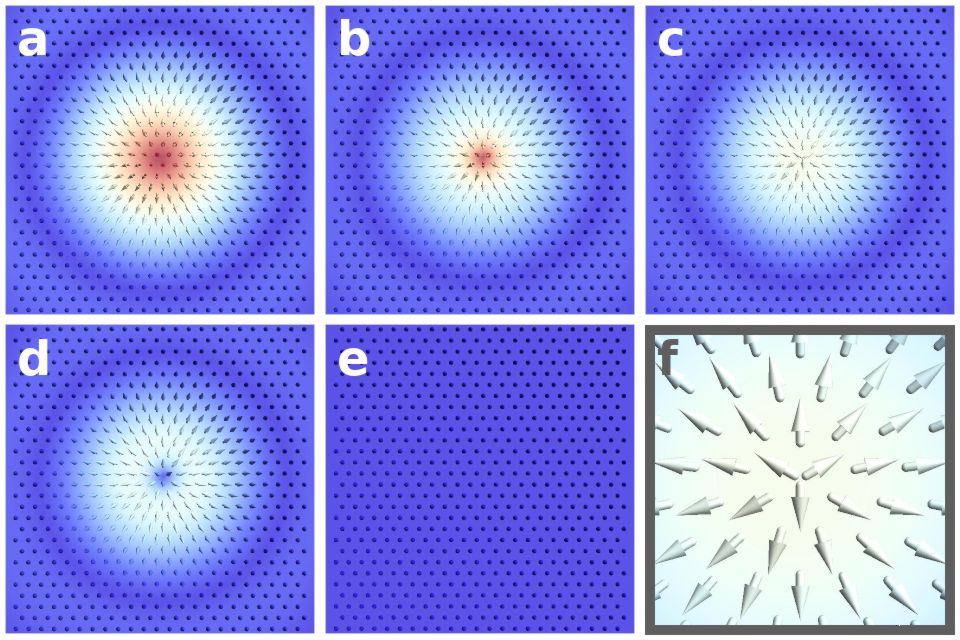}
    \vspace{1mm}

    \includegraphics[width=0.47 \textwidth]{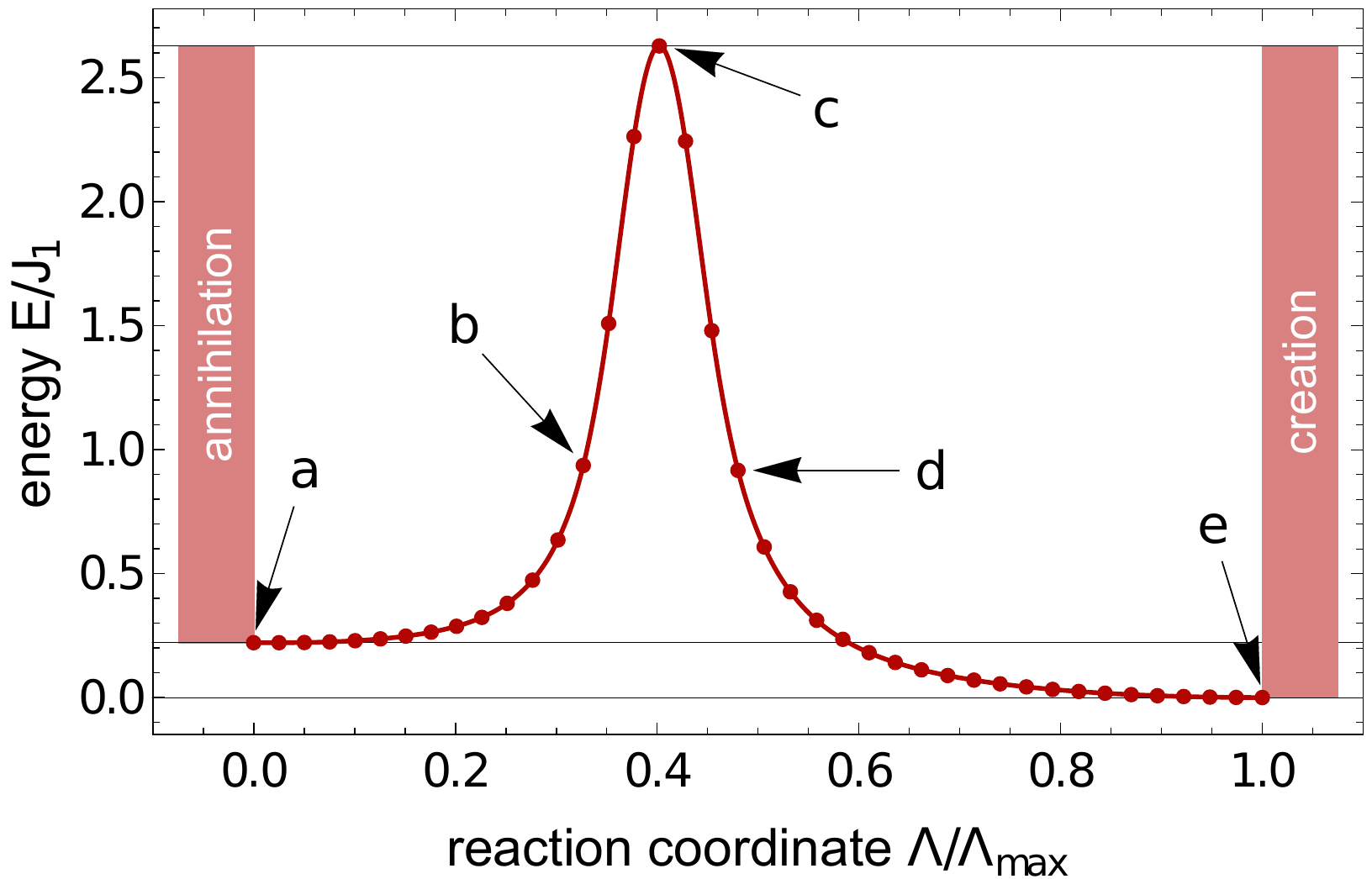}
    \caption{
        Minimal energy path for the creation/annihilation of a skyrmion in the symmetric model system, Eq.~\eqref{eq:frustration:model}, as obtained from the GNEB method, see Sec.~\ref{sec:appendix:minimalenergypath}.
        The upper panels (a-f) show the real-space magnetic texture in the proximity of the center of the skyrmion for various states along the minimal energy path.
        The color encodes the out-of-plane component of the magnetization.
        Panel (f) is a close-up of the saddle point texture in panel (c).
        The lower panel shows the energy $E/J_1$ along the minimal energy path as a function of the reaction coordinate $\Lambda$, see Sec.~\ref{sec:appendix:minimalenergypath}. 
        The energy is evaluated with respect to the polarized phase (e).
        The arrows indicate the position of the upper panels in the minimal energy path.
        These results were obtained for $J_2/J_1=31/90$, $\mu_s H/J_1=1/450$, and $K/J_1=1/900$.
        % scaleindex si=4
        % {s, J2, H, K} = {3, -(31/90), 1/450, 1/900}
        % see Mathematica-Notebook for more details.
    }
    \label{fig7}
\end{figure}

For the following analysis of the stability of skyrmions in symmetric magnets, we consider a minimal model as proposed in Ref.~[\onlinecite{leonov2015multiply}]:
Classical Heisenberg spins $\vec{m}_i = \vec{M}_i/M$ on a triangular lattice interact ferromagnetically with their nearest neighbors and antiferromagnetically with their next nearest neighbors.
Hence the system is frustrated as both interactions can never be optimized simultaneously and, as a compromise, co-planar spirals can form.
In addition, the external magnetic field $\vec{H}=H\hat{z}$ and a uni-axial anisotropy $K$ introduce a preferred axis for the magnetization and stabilize individual skyrmions.
The energy of the magnetic momenta $\vec{m}_i$ located on lattice sites $\vec{r}_i$ then reads
\begin{equation}
\begin{split}
    E =
    & - J_1 \sum_{\langle i, j \rangle} \vec{m}_i \!\cdot\! \vec{m}_j
    + J_2 \sum_{\langle\!\langle i, j \rangle\!\rangle} \vec{m}_i \!\cdot\! \vec{m}_j \\
    & - \mu_s H \sum_i m_i^z
    - K \sum_i (m_i^z)^2,
\end{split}
\label{eq:frustration:model}
\end{equation}
where we have adjusted the signs such that $J_1>0$ describes a ferromagnetic nearest-neighbor, while $J_2>0$ encodes the antiferromagnetic next-nearest-neighbor coupling.
The existence of metastable skyrmions requires\cite{leonov2015multiply} that $J_2 > \frac{1}{3} J_1$.

An example of a minimal energy path (MEP) obtained with the geodesic nudged elastic band method (GNEB) for the decay of a skyrmion into the polarized state is shown in Fig.~\ref{fig7} for $J_2/J_1=31/90$, $\mu_s H/J_1=1/450$, and $K/J_1=1/900$.
In contrast to the above discussed model with DMI, Eq.~\eqref{eq:dmi:model:discrete}, the energy functional of the centro-symmetric magnetic is symmetric under rotations of the magnetic momenta around the $\hat{z}$-axis, hence the helicity of a skyrmion is a zero mode and leads to interesting dynamics\cite{zhang2017skyrmion,leonov2017edge,lohani2019quantum}.
However, this additional degree of freedom does not influence the minimal energy path.
Thus, for convenience, the skyrmion in Fig.~\ref{fig5} is displayed with the same helicity as in Sec.~\ref{sec:dmi} but the result is the same for Bloch-type or any other helicity.

As can be clearly seen from the real-space images along the MEP, Figs~\ref{fig7}a-e, the GNEB method now finds a completely different decay mechanism compared to the DMI-stabilized skyrmion, Figs~\ref{fig1}a-e.
Instead of shrinking the entire skyrmion to the size of a lattice constant, the spins in the inner region with $m^z\leq0$ turn in-plane, forming a large vortex, while the outer region remains almost constant.
The saddle point, Figs~\ref{fig7}c or f, is assumed when the core of the skyrmion is approximately in-plane with a $120^\circ$ vortex-like singularity centered on a triangular plaquette.
The transition from the saddle point to the polarized state then occurs via first rotating the spins around the core of the vortex out of plane and then smoothly turning the remaining texture in the $\hat{z}$-direction.
The energy along the MEP is shown in the lower panel of Fig~\ref{fig7}.
The energy of the saddle point is an order of magnitude larger than the energy of the skyrmion, in contrast to the DMI-skyrmion, Fig.~\ref{fig1}, whose energy is of the same order as the energy of the saddle point.
However, a comparison in units of the nearest neighbor exchange interaction $J$ or $J_1$  reveals that the energy of the saddle point is, in fact, smaller than for the DMI-skyrmions. 

In the following we will analyze the energy barrier set by the saddle point and find that  the skyrmion and the saddle point are governed by different physics which leads to distinct energy scales for both.
This is in contrast to the previous section, where both the skyrmion and the saddle point are well described within a micromagnetic approximation.

%----------------------------------------------------------------------------------------
%----------------------------------------------------------------------------------------
\subsection{An effective continuum theory for large skyrmions}
\label{sec:frustration:largeskyrmions}
%----------------------------------------------------------------------------------------

In order to study the energy of the saddle point of the MEP for a large skyrmion, we can derive an effective energy functional with the same small-$q$ expansion as in Sec.~\ref{sec:dmi:largeskyrmions}.
In the absence of DMI, skyrmions are stabilized by a $(\nabla \vec{m})^2$ contribution to the energy with {\em negative} sign. To obtain a well-defined field theory bounded from below, one has to consider also the next order $\mathcal{O}(q^4)$ as already computed in Eq.~\eqref{eq:dmi:model:continuous:correction:triangle}.
The resulting $\mathcal{O}(q^4)$ continuum approximation of the energy functional reads
\begin{equation}
\begin{split}
    \delta E[\vec{m}]/J_1 = \!\int\!\!\mathrm{d}^2r \,\,
   &-\frac{\mathcal{I}_1}{2} (\nabla \vec{m})^2 
    + \frac{\mathcal{I}_2}{2} (\nabla^2 \vec{m})^2 \\
   &- \mu_0 \mathcal{H} (m_z-1) 
    - \mathcal{K} (m_z^2-1) \,\,.
\end{split}
\label{eq:frustration:model:continuous}
\end{equation}
Note that we use different conventions compared to the previous section. Energies are measured in units of $J_1$ and the coupling constants are obtained as
\begin{equation}
    \mathcal{I}_1 = \sqrt{3}\left(3 \frac{J_2}{J_1}-1\right), \quad 
    \mathcal{I}_2 = \frac{a 
    ^2\sqrt{3}}{16}\left(9 \frac{J_2}{J_1}-1\right),
\label{eq:frustration:model:continuous:units}
\end{equation} 
while $\mathcal{H}=2H/(\sqrt{3}a^2 J_1)$ and $\mathcal{K}=2K/(\sqrt{3}a^2 J_1)$.
Importantly, skyrmions are only stabilized~\cite{leonov2015multiply} for \textit{negative} spin stiffness $-J_1 \mathcal{I}_1$, i.e., for $\mathcal{I}_1>0$ or  $J_2>\frac{1}{3}J_1$. 
In this regime, we obtain $\mathcal{I}_2>0$.
Both $\mathcal{I}_1$ and $\mathcal{I}_2$ depend only on $J_2/J_1$ and are related by $\mathcal{I}_2=a^2 (\sqrt{3}/8 + 3\,\mathcal{I}_1/16)$.

From a dimensional analysis we can extract the length scale $\xi$
\begin{equation}
    \xi = \sqrt{\frac{\mathcal{I}_2}{\mathcal{I}_1}} = \frac{a}{4} \,\sqrt{\frac{J_1 - 9 J_2}{J_1 - 3J_2}} %\sim \left(1 - 3\frac{J_2}{J_1}\right)^{-\frac{1}{2}}
\label{eq:frustration:model:continuous:lengthscale}
\end{equation}
which is tuned to vary the size of the skyrmion.
Large skyrmions with $\xi\gg a$ are stabilized for small frustration $\mathcal{I}_1\to0$, i.e., by fine-tuning the nearest and next-nearest neighbor couplings $J_2\to \frac{1}{3}J_1$.
Experimentally, this is more complicated to achieve than large DMI-stabilized skyrmions, as the latter only require a small spin-orbit coupling, realized in most materials.

\begin{figure}
    \center
    \includegraphics[width=0.47 \textwidth]{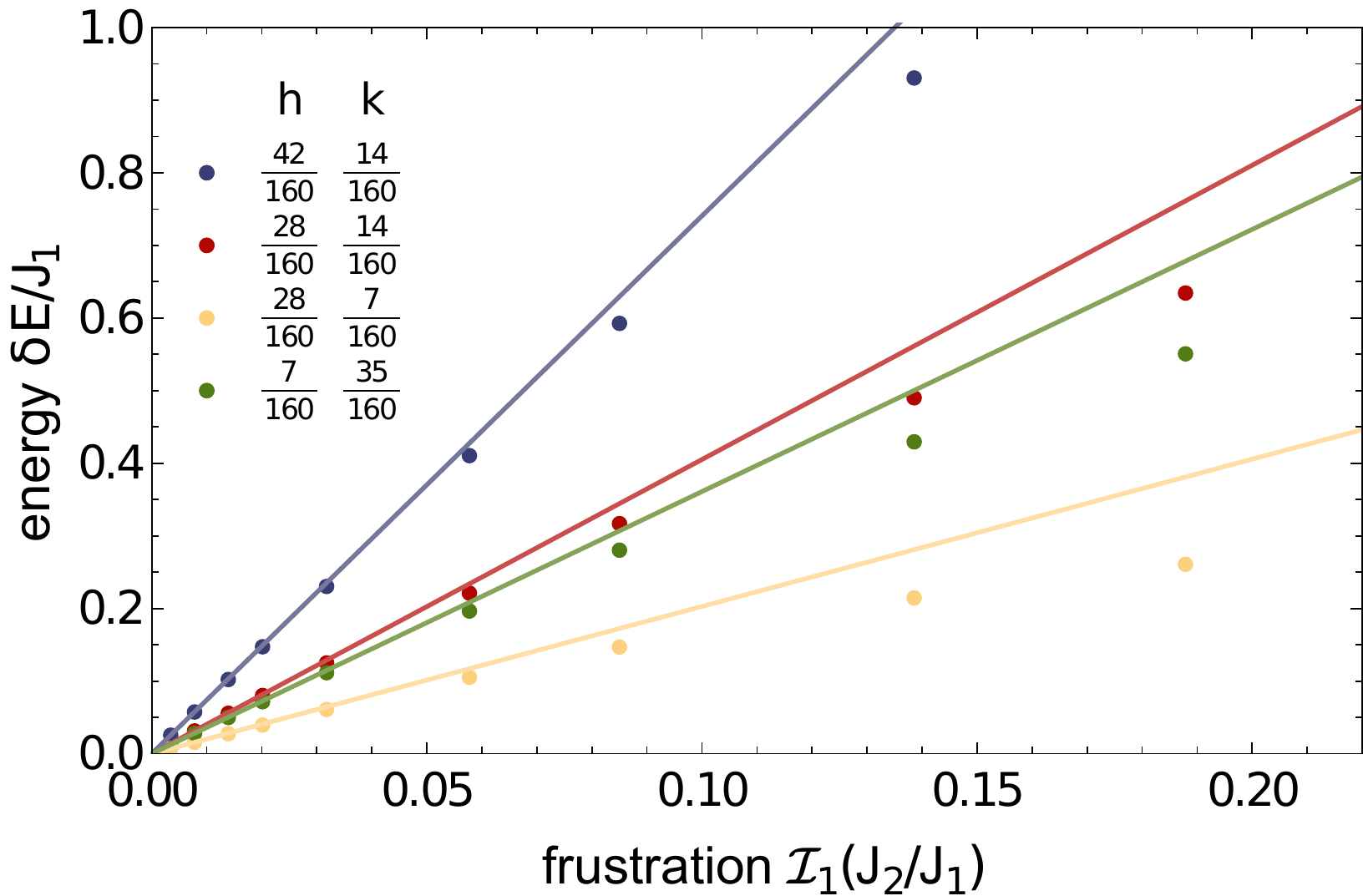}
    \caption{
        Energy of an isolated skyrmion relative to the polarized state.
        The plot shows the results obtained for the model Eq.~\eqref{eq:frustration:model} on a triangular lattice.
        The magnetic field $\mu_0 H/J_1$ and uni-axial anisotropy $K/J_1$ is chosen such that dots of the same color map to the same rescaled continuum field $h$ and anisotropy $\kappa$, but are evaluated for different values of $J_2/J_1$.
        The data is plotted as a function of the resulting frustration $\mathcal{I}_1(J_2/J_1)$, see Eq.~\eqref{eq:frustration:model:continuous:units}.
        The solid lines are fits for the expected linear behavior at vanishing frustration, see Eq.~\eqref{eq:frustration:model:continuous:rescaled:limit}, using only the point with lowest frustration for fitting.
    }
    \label{fig8}
\end{figure}

A rescaling of the energy of the skyrmion using dimensionless units with $\vec{r} = \xi\,\tilde{\vec{r}}$ gives
\begin{equation}
\begin{split}
    \delta E/J_1 = \mathcal{I}_1 \!\int\!\!\mathrm{d}^2\tilde{r} \,\,
   &- \frac{1}{2} (\tilde{\nabla} \vec{n})^2 
    + \frac{1}{2} (\tilde{\nabla}^2 \vec{n})^2 \\
   &- h (n_z-1) 
    - \kappa (n_z^2-1) \,\,,
\end{split}
\label{eq:frustration:model:continuous:rescaled}
\end{equation}
with $h=\mu_0\mathcal{H}\,\mathcal{I}_2/\mathcal{I}_1^2$ and $\kappa=\mathcal{K}\,\mathcal{I}_2/\mathcal{I}_1^2$.

For a fixed value of the rescaled parameters ($h$,$\kappa$) the energy of a skyrmion is therefore linear in $\mathcal{I}_1\sim(a/\xi)^{2}$ and vanishes for larger and larger skyrmions, 
\begin{equation}
    \delta E/J_1 = \mathcal{I}_1 C(h,\kappa) \,\,.
\label{eq:frustration:model:continuous:rescaled:limit}
\end{equation}
where $ C(h,\kappa)$ is a numerical constant depending on the parameters $h$ and $\kappa$.
This linear dependence on $\mathcal{I}_1$ is confirmed in Fig.~\ref{fig8} where we plot the energy of the skyrmion in the atomistic model, Eq.~\eqref{eq:frustration:model}, as function of the frustration for various fixed values of the rescaled external field $h$ and rescaled anisotropy $\kappa$.

\subsubsection{The saddle point of the minimal energy path for large skyrmions}
\label{sec:frustration:largeskyrmions:saddlepoint}

In Sec.~\ref{sec:dmi:largeskyrmions:saddlepoint}, we have argued that the DMI-stabilized skyrmion has to shrink before it is destroyed along its minimal energy path because a vortex of radius $R$ is associated with an energy cost of $\mathcal{J} \pi \ln(R/a)$ which can be minimized by reducing the size of $R$. For a skyrmion stabilized by frustration, the positiv constant $\mathcal{J}$ is replaced by the negative constant $-J_1 \mathcal{I}_1$. Therefore there is no reason to expect a shrinking of the skyrmion. Instead, the decay occurs via the formation of a vortex with a size comparable to the skyrmion size, fully consistent with our numerical results, see Fig.~\ref{fig7}. Fig.~\ref{fig7}c and Fig.~\ref{fig7}f show that the saddle point configuration is indeed a vortex state.

To estimate the energy cost of a vortex, we  decompose the skyrmion into three parts:
(i) an outer region $\tilde{r} > \tilde R$ where the magnetization smoothly turns into the polarized state,
(ii) the region close to the center $\tilde{r}_0 < \tilde{r} < \tilde{R}$ where the magnetization is approximately in-plane and pointing outwards $\vec{n}(\tilde{\vec{r}}) = \hat{r}$ and
(iii) the core of the vortex $\tilde{r} < \tilde{r}_0 = r_0/\xi $, where $r_0$ is a length scale of the order of the lattice constant, $r_0\sim a$. 
In region (iii), one has to take into account the discrete lattice and the microscopic interactions as the singular nature of the vortex impedes a continuum description. 
Its energy $e^{(iii)}(J_1,J_2)$ is mainly determined by the large interactions $J_1$ and $J_2$ and only weakly affected by the small anisotropy or external field.
In turn, the magnetization in the regimes (i) and (ii) varies smoothly and hence we can use the continuum model, Eq.~\eqref{eq:frustration:model:continuous:rescaled}.
From the outer region (i) we obtain a term 
$\mathcal{I}_1 C^{(i)}(h,\kappa)$ which is linear in the small prefactor $\mathcal{I}_1$  and depends on $h$ and $\kappa$. More interesting is region (ii) as the gradient terms give contributions which are singular as a function of $r_0$. Collecting all terms we find
\begin{align}
    \frac{\delta E}{J_1} &\approx 
    \mathcal{I}_1 C^{(i)}(h,\kappa)
    - \mathcal{I}_1 \pi \!\int_{\tilde{r}_0}^{\tilde{R}} \! \left(\frac{1}{\tilde{r}} - \frac{1}{\tilde{r}^3}\right)\mathrm{d}\tilde{r} \nonumber \\
  &\qquad +\mathcal{I}_1 C^{(ii)}(h,\kappa)
    + e^{(iii)}(J_1,J_2) \nonumber \\
    %&= 
    %\mathcal{I}_1 C^{(i)}
    %+\mathcal{I}_1 \pi \left( - \ln\left(\!\frac{\tilde{R}}{\tilde{r}_0}\!\right)
    %+ \left(\!\frac{1}{2\tilde{r}_0^2} - \frac{1}{2\tilde{R}^2}\!\right) \right)  
    %+ E^\text{v}(J_1,J_2)\\
    %&= 
    %\mathcal{I}_1 C_1
    %+ \mathcal{I}_1 \pi \ln\left( r_0/\xi \right) 
    %+ \frac{\mathcal{I}_2 \pi}{ 2 r_0^2}   
    %+ E^\text{v}(J_1,J_2)\\
    %&= 
    %\mathcal{I}_1 C_1
    %+ \frac{\mathcal{I}_1 \pi}{2} \ln\left( r^2_0/\xi^2 \right) 
    %+ \frac{\sqrt{3} \pi}{ 16 r_0^2}   
    %+ E^\text{v}(J_1,J_2)\\
    %&= 
    %\mathcal{I}_1 C_1
    %+ \frac{\mathcal{I}_1 \pi}{2} \ln\left( \frac{ \mathcal{I}_1 r^2_0}{\mathcal{I}_2} \right) 
    %+ \frac{\sqrt{3} \,\pi}{ 16 \, r_0^2}   
    %+ E^\text{v}(J_1,J_2)\\
    %&= 
    %\mathcal{I}_1 C_1
    %+ \frac{\mathcal{I}_1 \pi}{2} \ln\left( \frac{ \mathcal{I}_1 r^2_0}{\mathcal{I}_2} \right) 
    %+ C_0\\
    &= 
    e_0-\frac{\pi}{2} \mathcal{I}_1  \ln\!\left(\frac{1}{\mathcal{I}_1}  \right) 
    +  \mathcal{I}_1  e_1(h,\kappa)
\label{eq:frustration:largeskyrmions:saddlepoint:vortexenergy}
\end{align}
where $e_0 = \frac{\sqrt{3} \,\pi}{ 16 \, \tilde{r}_0^2} + e^{(iii)}(J_1,J_2)$ is a constant independent of $\mathcal{I}_1$ in the limit of small $\mathcal{I}_1$. 
$e_0 J_1$ is the energy of the saddle point in the limit $\mathcal{I}_1=0$. 
It cannot be calculated in any continuum theory as it captures the singular center of the vortex but, instead, it can be computed from from the atomistic energy functional Eq.~\eqref{eq:frustration:model} by considering a vortex centered on a triangular plaquette for $\mathcal{I}_1=0$, i.e., for $J_2=J_1/3$ and $H=K=0$. Note that for these parameters the energy of an infinite-size vortex is finite (as $\mathcal{I}_1$ vanishes), while its energy diverges for $J_2 \neq J_1/3$. A very good quantitative estimate for $e_0$ with a precision of about $1\%$ can simply be obtained by calculating the energy of the in-plane vortex configuration $\vec m_i=\frac{\vec r_i - \vec r_c}{|\vec r_i -\vec r_c|}$, where $r_c$ is the center of a triangular plaquette. From this calculalation we obtain $e_0=2.8482$.
A simple optimization of this solution obtained by minimizing the energy while keeping the spins in-plane yields with very high precision the value
\begin{equation}
e_0 \approx 2.8156 \label{eq:e0}
\end{equation}
for classical spins on a triangular lattice.

\begin{figure}
    \center
    \includegraphics[width=0.47 \textwidth]{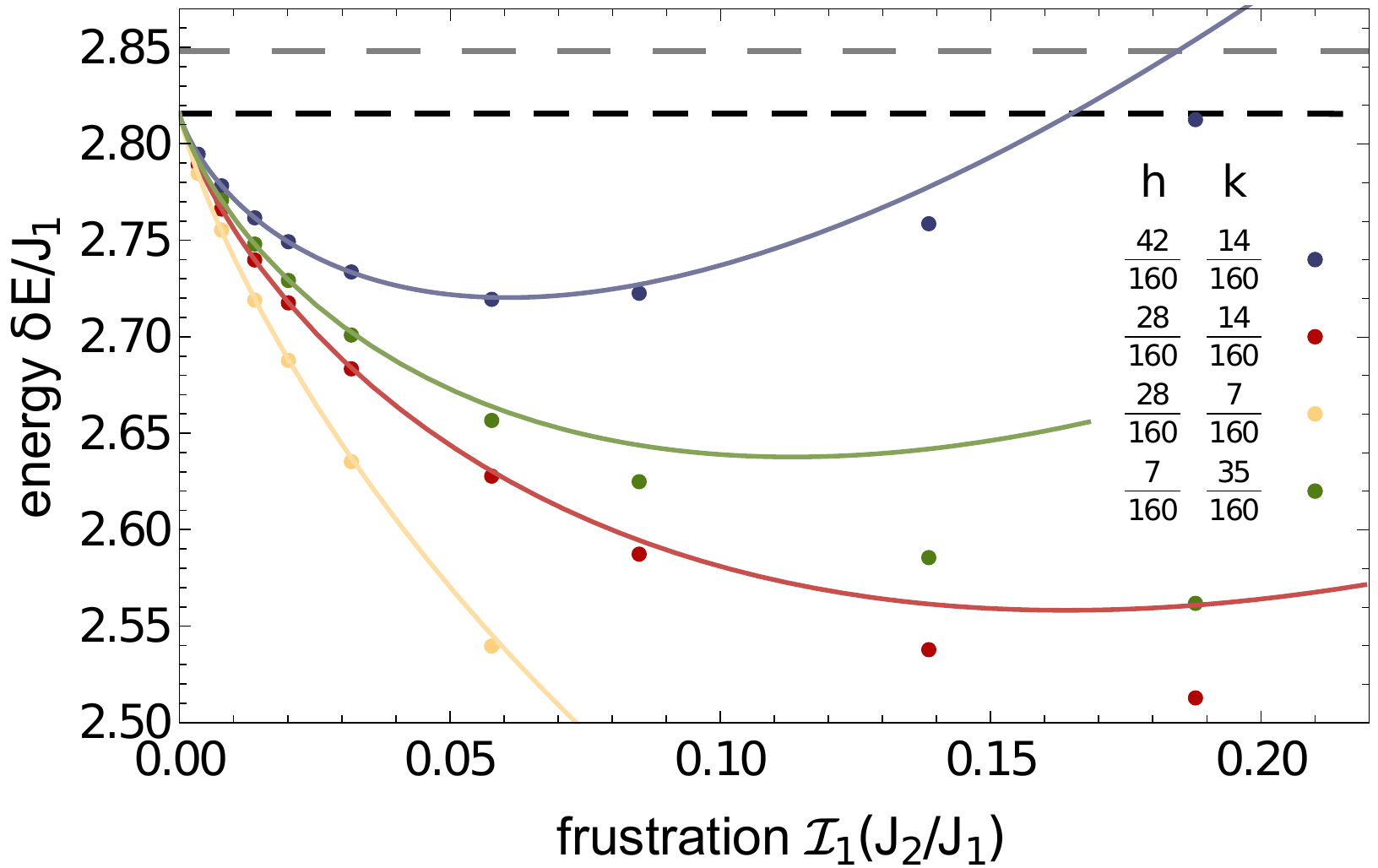}
    \caption{
        Saddle-point energy of large skyrmions (see Fig.~\ref{fig7}) in a frustrated magnet on a triangular lattice, Eq.~\eqref{eq:dmi:model:discrete}.
        The magnetic field $\mu_0 H/J_1$ and uni-axial anisotropy $K/J_1$ are chosen to obtain the fixed values of $h$ and $\kappa$ shown in the legend of the figure.
        The data is plotted as a function of the resulting frustration $\mathcal{I}_1(J_2/J_1)$, see Eq.~\eqref{eq:frustration:model:continuous:units}.
        The solid lines are fits using Eq.~\eqref{eq:frustration:largeskyrmions:saddlepoint:vortexenergy} with $e_1(h,\kappa)$ as the only fitting parameter ($e_1 = 2.84$ (blue), $1.27$ (red), $-0.20$ (yellow), and $1.85$ (green)).
        The black dashed line denotes the asymptotic value of the energy for $\mathcal{I}_1 \to 0$, Eq.~\eqref{eq:e0} (gray dashed line: approximate value obtained for a unrelaxed, symmetric vortex, $e_0 \approx 2.8482$, see text).  
    }
    \label{fig9}
\end{figure}

In Fig.~\ref{fig9} we compare the analytical formula \eqref{eq:frustration:largeskyrmions:saddlepoint:vortexenergy}
to GNEB calculations of the energy barrier for various sets of parameters. In the large-skyrmion limit, $\mathcal{I}_1 \to 0$, we recover the result $e_0 J_1$ from Eq.~\eqref{eq:e0} with high precision. To describe the deviations from this asymptotic value, we can use Eq.~\eqref{eq:frustration:largeskyrmions:saddlepoint:vortexenergy}, where we treat $e_1(h,\kappa)$ for fixed $h$ and $\kappa$ as the only fitting parameter. Here the logarithmic corrections characteristic of the vortex solution are crucial for the energy of the saddle point for finite but small $\mathcal{I}_1$, see Eq.~\eqref{eq:frustration:largeskyrmions:saddlepoint:vortexenergy}. We thus obtain a highly accurate quantitative description of the energy barrier for $\mathcal{I}_1 \lesssim 0.1$, fully corrobating the physical picture of a large-vortex collapse path for (large) skyrmions stabilized by frustrating interactions.

Our analytical treatment was originally based on the picture of a decay path of the skyrmion which is rotationally symmetric.
As discussed in more detail in Appendix~\ref{sec:appendix:asymmetricdecay}, for large anisotropies the decay path of the skyrmion obtained from GNEB calculations is highly asymmetric \cite{meyer2019isolated} as it is more favourable to locate the vortex in a region where skyrmion-spins are in-plane. As the transition still occurs via a large vortex, this does, however, not invalidate our analysis and Eq.~\eqref{eq:frustration:largeskyrmions:saddlepoint:vortexenergy} is still valid for small $\mathcal{I}_1$. Only  the numerical value of the constant $e_1(h,\kappa)$ is affected by the asymmetric decay path.

%----------------------------------------------------------------------------------------
%----------------------------------------------------------------------------------------
\subsubsection{Decay into an anti-skyrmion}
\label{sec:frustration:decaytoantiskyrmion}
%----------------------------------------------------------------------------------------

\begin{figure}
    \center
    \includegraphics[width=0.47 \textwidth]{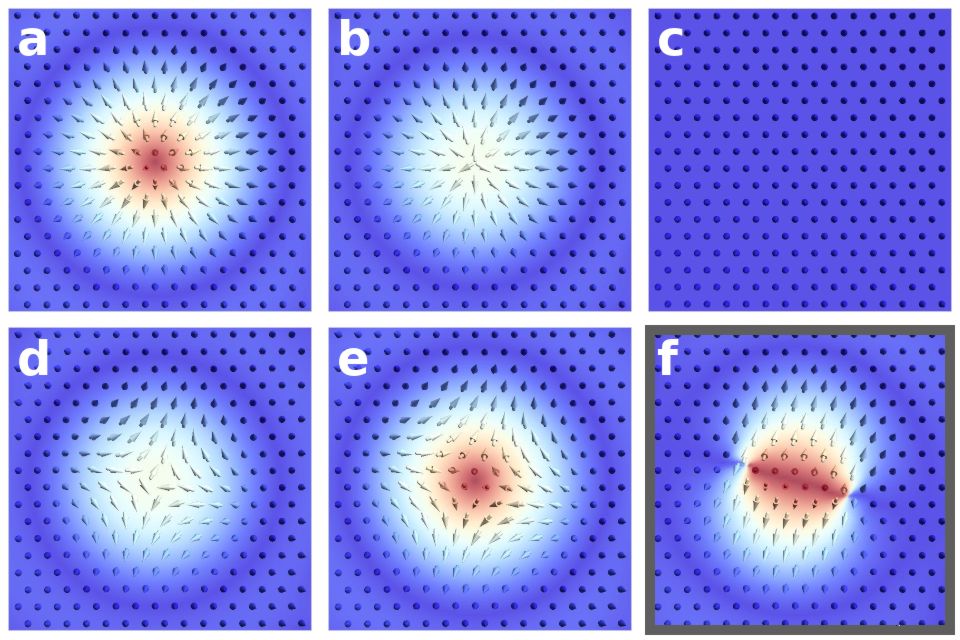}
    \vspace{1mm}

    \includegraphics[width=0.47 \textwidth]{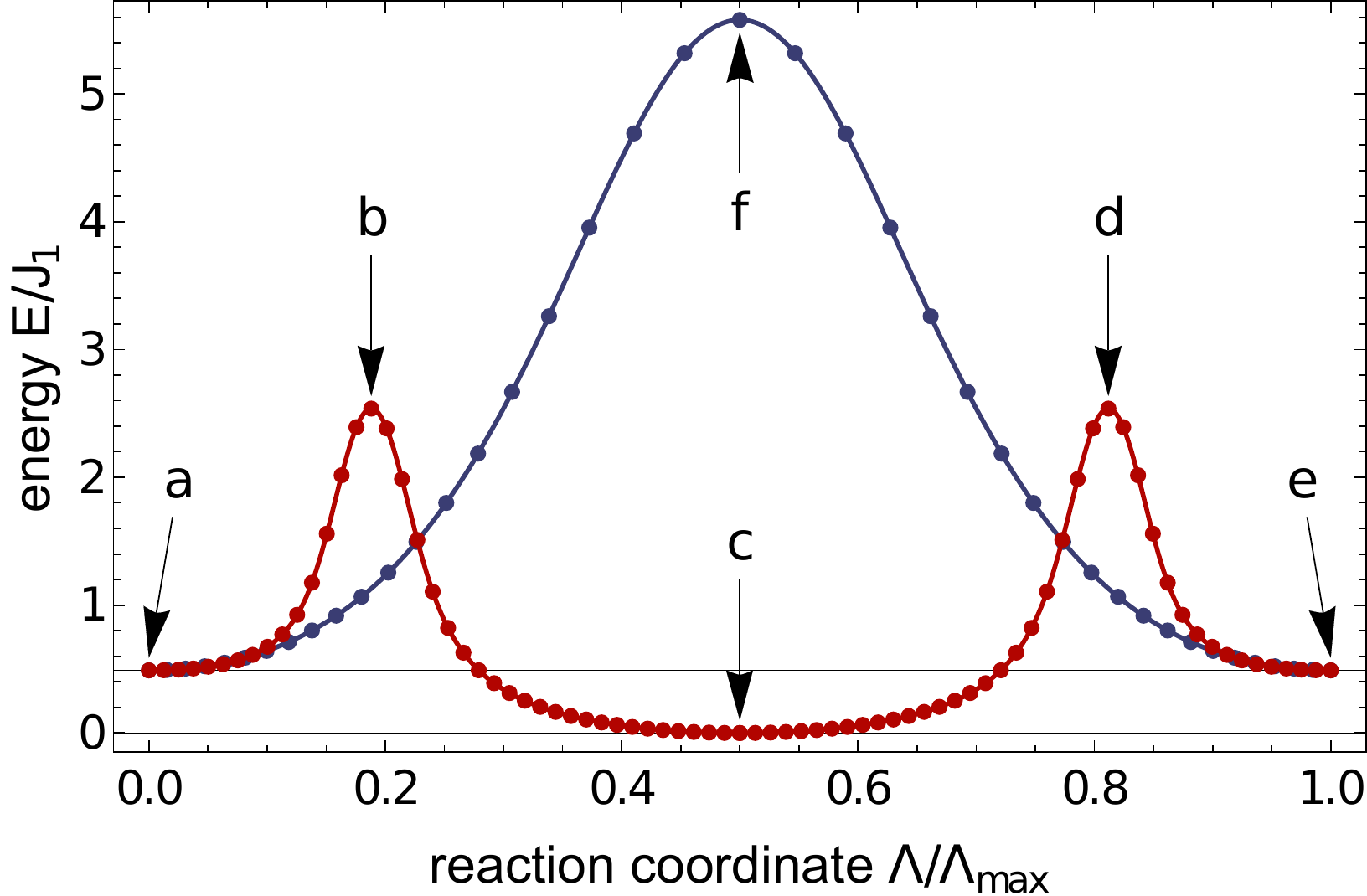}
    \caption{
        Minimal energy path for the transition of a skyrmion into an anti-skyrmion as obtained from the GNEB method, see Sec.~\ref{sec:appendix:minimalenergypath}.
        The upper panels (a-e) show the real-space magnetic texture in the proximity of the center of the (anti-)skyrmion for various states along the minimal energy path (lower panel, red data).
        The color encodes the out-of-plane component of the magnetization.
        Panel (f) shows the maximum of the initialized path (lower panel, blue data) with two vortices which is unstable.
        These results are obtained for $J_2/J_1=9/25$, $\mu_s H/J_1=3/250$, $K/J_1=3/500$ in model Eq.~\eqref{eq:frustration:model}.
        %original data: {s, J2, H, K} = {2, -(9/25), 3/250, 3/500}
    }
    \label{fig10}
\end{figure}

The decay of a skyrmion into a polarized state is not the only decay mechanism which can be of importance.
Other mechanisms include the bimeron-instability\cite{ezawa2011compact} where the skyrmion elongates arbitrarily which can trigger the duplication of a skyrmion\cite{muller2018duplication}.
For large uni-axial anisotropy when the decay is asymmetric, see Sec.~\ref{sec:appendix:asymmetricdecay}, we also observe an intermediate skyrmion-like state with zero winding number similar to Ref.~\citenum{meyer2019isolated} which, however, decays immediately.
In the frustrated system without DMI discussed here, the lack of a handedness leads to the degeneracy of skyrmions and anti-skyrmions.
Furthermore, they not only have the same energy but also the same magnetization which enables quantum mechanical tunnelling\cite{lohani2019quantum}.

For the calculation of the minimal energy path from a skyrmion to an anti-skyrmion we prepare an initial path which exhibits a maximum with two simultaneous vortices, see Fig.~\ref{fig10}f.
This artificial maximum has approximately twice the energy of a single singularity.
During the application of the GNEB method, see Sec.~\ref{sec:appendix:minimalenergypath}, the initial maximum decays into two separate maxima with one vortex each.
The intermediate minimum turns out to be the fully polarized state.
Therefore, the total minimal energy path is the decay of the skyrmion into a ferromagnetic state followed by the creation of the anti-skyrmion in a reverse process.
The resulting path is shown in Fig.~\ref{fig10} (red) for $J_2/J_1=9/25$, $\mu_s H/J_1=3/250$, $K/J_1=3/500$ and can be constructed from the individual decay/creation paths discussed in Sec.~\ref{sec:frustration:largeskyrmions}.

Since the minimal energy path is via an intermediate polarized state, we conclude that a direct transition from a skyrmion to a antiskymrion by thermal fluctuations is very unlikely and of no practical importance. 
Assuming that skyrmion and antiskyrmion have an energy which is smaller than the energy of the ferromagnet, it is more likely that skyrmions (or antiskyrmions) proliferate spontaneously.
This conclusion holds at least in the absence of defects and boundaries.

%----------------------------------------------------------------------------------------
%----------------------------------------------------------------------------------------
\subsubsection{Conservation of magnetization}
\label{sec:frustration:conservation}
%----------------------------------------------------------------------------------------

The Hamiltonian of Eq.~\eqref{eq:frustration:model} describes formally a system where the total magnetization in z-direction $M_z = \sum_\vec{r} m_z(\vec{r})$ is a conserved quantity. In a real material, the magnetization is never exactly conserved due to (weak) spin-orbit coupling or due to spin-lattice relaxation processes involving spin-flip terms and phonons. In numerical simulations of the Landau-Lifshitz-Gilbert equation, this is usually taken into account in a phenomenological way by introducing the Gilbert damping $\alpha$. 
The presence of spin-flip processes in the thermal fluctuations justifies that we have ignored spin conservation when calculating the  the relevant energy barriers.

There are, however, situations in which one would like to know the effective energy barrier in the presence of the magnetization constraint. This barrier will, for example, be important for the quantum-mechanical tunneling process from a skyrmion to an antiskyrmion in a frustrated magnet with little or no spin-orbit interactions\cite{lohani2019quantum}. 
Even in the classical limit, there can be the situation where a process, which does not requiring spin-relaxation and has a larger energy barrier, competes with a process with a lower barrier requiring spin relaxation.

 In general, any constraint to a system leads to an increase of its ground-state energy. Similarly, the energy of the saddle point, $\delta E |_{M_z^S}$, where the magnetization is fixed to the magnetization $M_z^S$ of the skyrmion, will be higher than the saddle point energy, $\delta E^\text{\tiny{SP}}$, without that constraint, $\delta E^\text{\tiny{SP}} \leq \delta E |_{M_z^S}$.
The vortex-style saddle point has a higher magnetization than the initial skyrmion state, $\Delta M_z =M_z^S- M_z^\text{\tiny{SP}}<0$.
Asumming an infinitely large system, we obtain an upper bound for the energy difference of the saddle points,  $\Delta E |_{M_z^0}- \delta E^\text{\tiny{SP}}$ by using a spin configuration which is identical to the saddle point configuration close to the origin, but it absorbs the excess magnetization $\Delta M_z$  by slightly tilting the ferromagnetic magnetization far away from the origin. This costs the energy $-(\mu_s H + 2 K) \Delta M_z$ and we therefore find
\begin{equation}
   0 \leq \delta E |_{M_z^S}-   \delta E^\text{\tiny{SP}} \! \leq -(\mu_s H + 2 K) \Delta M_z \sim J_1 (h +\kappa) \mathcal{I}_1,
\end{equation}
where we used that $|\Delta M_z| \sim  \xi^2 \sim a^2/\mathcal{I}_1$. At least in the limit of small $\mathcal{I}_1$,
the energy barrier $\delta E |_{M_z^S}$ for processes which conserve magnetization is therefore only slightly higher than the barriers $\delta E^\text{\tiny{SP}}$ calculated in Fig.~\ref{fig9}.

\section{Discussion}
\label{sec:discussion}
%----------------------------------------------------------------------------------------

We have calculated numerically the barriers for annihilation and creation of successively larger skyrmions in monolayers of chiral magnets and frustrated magnets. Three different scenarios for the destruction of a skyrmion state emerged. DMI-stabilized skyrmions in chiral magnets decay by shrinking. Depending on the sign of the 4th-order derivative terms, $\bar{\mathcal K}_4$, the saddle-point spin configuration which determines the energy barrier is either smooth (i) or singular (ii) on the length scale set by the lattice constant. 

Most common seems to be case (i),  $\bar{\mathcal K}_4<0$, where the saddle-point energy of a large DMI skyrmion is simply given by $4 \pi \mathcal{J}$, the energy of the famous Belavin-Polyakov skyrmion in a system with spin-stiffness $\mathcal{J}$.  Corrections proportional to  $(D/J)^{2/3}$ always {\em reduce} the saddle point energy. 
In this case, one can use micromagnetic simulations to calculate the energy barrier but to obtain the correction to the universal value $4 \pi \mathcal{J}$, one has to include 4th-order derivative terms in the model, see Appendix~\ref{sec:appendix:continuumlimit}. 
While the spin stiffness $\mathcal J$ can be measured directly using Neutron scattering and is routinely calculated using ab-initio methods, a measurement (or ab-initio calculation) of $\bar{\mathcal K}_4$ is needed as an input to the micromagnetic simulations, which is often much more challenging to obtain.

In case (ii), $\bar{\mathcal K}_4>0$, the saddle point energy of a DMI skyrmion is determined by microscopic physics on the length scale of the lattice spacing. The saddle point energy is typically larger than $4 \pi \mathcal{J}$ in this case and determined by a spin configuration where spin orientation varies rapidly on the length scale of a lattice constant. This case is, for example, realized when frustrating interactions reduce the size of the spin-stiffness $\mathcal{J}$ while keeping the energy of a local vortex configuration high. While one can treat such a situation easily by brute-force numerics in idealized classical-spin models, it is very difficult to model this situation for an insulating spin-$1/2$ quantum spin system or for metallic compounds. In the latter case it would be interesting to combine the GNEB approach with an electronic ab-initio calculation.

A third case is realized in Heisenberg ferromagnets where frustration induces a {\em negative} spin stiffness, which provides an alternative stabilization mechanism for skyrmions. In this case, the energy barrier is determined by a vortex solution which has a size of the order of the skyrmion radius. Here the main contribution to the energy arises from the singular core of the vortex state, which again depends on microscopic details on the length scale of the lattice spacing. For classical spin models and large skyrmions, it was possible to get an accurate value for the energy barrier by simply considering an infinite vortex configuration in a model fine-tuned to have a vanishing spin-stiffness.

In our study, we have only considered a magnetic monolayer. It would be very interesting to extend the analysis to 
multi-layer compounds or even three dimensional systems. For very thin systems and a few Heisenberg-coupled 
monolayers, we can expect that one can simply multiply all energies by the number of layers. For thick system, in contrast, the nature of the saddle point configuration changes. The relevant singular spin configuration is in this case a Bloch point (or a pair of Bloch points)\cite{milde2013unwinding,schuette2014dynamics,rybakov2015newtype,wild2017entropy}. 
In this case, the core-contribution of the energy of order of the exchange coupling $J$ is expected to be much smaller (see, e.g., supplement of Ref.~[\citenum{wild2017entropy}]) than contributions from the surrounding of order $J^2/D$.

In conclusion, we have studied the topological protection of two-dimensional skyrmions. For a large class of 2d systems, we could identify the precise conditions required to obtain a universal energy barrier approximately given by $4 \pi \mathcal{J}$. For other cases, however, microscopic details become more important which make the computation of the topological energy barrier a challenging task for real materials.

%----------------------------------------------------------------------------------------
%----------------------------------------------------------------------------------------
%----------------------------------------------------------------------------------------
\section{Acknowledgments}
\label{sec:acknowledgments}
%----------------------------------------------------------------------------------------
J.M. thanks G. P. M\"uller, P. Bessarab, N. Kiselev, M. Hoffmann, S. von Malottki, S. Meyer, and S. Bl\"ugel for the fruitful discussions.
This work was supported by the Deutsche Forschungsgemeinschaft (DFG, German Research Foundation) CRC 1238 project C04, project number 277146847.
We furthermore thank the Regional Computing Center of the University of Cologne (RRZK) for providing computing time on the DFG-funded   High Performance Computing (HPC) system CHEOPS as well as support.

\appendix

%----------------------------------------------------------------------------------------
\section{Parameters}
\label{sec:appendix:parameters}
%----------------------------------------------------------------------------------------

We chose the interaction parameters in our calculations by first fixing the parameters of the continuum theory, which allows to compare data sets described by the same (lowest order) continuum theory.

\subsection{Non-symmetric magnets}
For non-symmetric magnets with DMI, Sec.~\ref{sec:dmi}, we can map the atomistic model, Eq.~\eqref{eq:dmi:model:discrete}, onto the continuum model in dimensionless units, Eq.~\eqref{eq:dmi:model:continuous:rescaled}.
The length scale $\xi$ of the system fixes the DMI constant $D$:
\begin{equation}
 D/J = \frac{a}{\xi} \,\,.
\end{equation}
The dimensionless magnetic field $h=\mu_0 \mathcal{J}\mathcal{H}/\mathcal{D}^2$ and uni-axial anisotropy $\kappa=\mathcal{J}\mathcal{K}/\mathcal{D}^2$ are fixed for a series of parameters.
Hence, the atomistic interaction parameters used in the simulations are 
\begin{equation}
\mu_0 H/J = \frac{a^2}{\xi^2}\,h,\,\, K/J =\frac{a^2}{\xi^2}\,\kappa
\end{equation}
on the square lattice and
\begin{equation}
\mu_0 H/J = \frac{3\,a^2}{2\,\xi^2}\,h,\,\, K/J = \frac{3\,a^2}{2\,\xi^2}\,\kappa
\end{equation}
on the triangular lattice.
The data shown in Sec.~\ref{sec:dmi} is computed for the dimensionless parameters $(h,\kappa)=(1.1,0)$ (dark blue), $(1,0)$ (blue and red), $(0.75,0)$ (light blue), and $(0,1.3)$ (yellow).
The scales $\xi$ were chosen as $\xi/a=s/0.8$ with $s=1,2,4,8,16,32,64$, except for the anisotropy-stabilized skyrmion where the data for $s=64$ did not converge in a reasonable time.

\subsection{Centro-symmetric magnets}
For symmetric magnets, Sec.~\ref{sec:frustration}, the atomistic model, Eq.~\eqref{eq:frustration:model}, maps onto the continuum model in dimensionless units in Eq.~\eqref{eq:frustration:model:continuous:rescaled}.
Here, the length scale $\xi$ is given in Eq.~\eqref{eq:frustration:model:continuous:lengthscale} which yields an expression for the next-nearest neighbor interaction $J_2$:
\begin{equation}
 J_2/J_1 = \frac{1 - 16 (\xi/a)^2}{3 (3 - 16 (\xi/a)^2)} \,\,.
\end{equation}
For fixed dimensionless field $h=\mu_0\mathcal{H}\,\mathcal{I}_2/\mathcal{I}_1^2$ and anisotropy $\kappa=\mathcal{K}\,\mathcal{I}_2/\mathcal{I}_1^2$, the corresponding atomistic parameters then read
\begin{equation}
\begin{split}
 \mu_0 H/J_1 &= \frac{3 \,h}{(\xi/a)^2 (-3 + 16 (\xi/a)^2)} ,\\
 K/J_1 &= \frac{3 \,\kappa}{(\xi/a)^2 (-3 + 16 (\xi/a)^2)}.
 \end{split}
\end{equation}
The data shown in Sec.~\ref{sec:frustration} is computed for the dimensionless parameters $(h,\kappa)=(\frac{21}{80},\frac{7}{80})$, $(\frac{7}{40},\frac{7}{80})$, $(\frac{7}{40},\frac{7}{160})$, and $(\frac{7}{160},\frac{35}{160})$ as indicated in the plots.
The scales $\xi$ were chosen as $\xi/a=\frac{\sqrt{7}}{4}s$ with $s=1\frac{3}{4}, 2, 2\frac{1}{2}, 3, 4, 5, 6, 8, 12$ with the exception of $(h,\kappa)=(\frac{7}{160},\frac{35}{160})$, where $s=12$ did not converge in a reasonable time.
Note, that the frustration parameter scales as 
\begin{equation}
 \mathcal{I}_1 = \frac{2\sqrt{3}}{-3 + 16(\xi/a)^2} \sim (a/\xi)^{2}\,\,,
\end{equation}
therefore, the data in Sec.~\ref{sec:frustration} is not plotted as a function of the inverse size of the skyrmion, in constrast to the data presented in Sec.~\ref{sec:dmi}.

%----------------------------------------------------------------------------------------
\section{Methods}
\label{sec:appendix:methods}
%----------------------------------------------------------------------------------------

%----------------------------------------------------------------------------------------
\subsection{Methods: Minimal energy path calculations}
\label{sec:appendix:minimalenergypath}
%----------------------------------------------------------------------------------------
For the calculation of minimal energy transition paths we use the geodesic nudged elastic band (GNEB) method with additional climbing image (CI) as described in Ref.~\citenum{bessarab2015method}.
In some figures, however, we use an advanced method for the calculation of the spring forces between adjacent images as explained below.

We first prepare a single skyrmion in a polarized background with periodic boundary conditions and relax the texture into a local energy minimum.
The sample size is chosen large enough so that the self-interaction effects over the periodic boundaries are negligible.
An interpolation between this skyrmion and the polarized state (or the anti-skyrmion) over a total of $40$ layers serves as the starting point for the GNEB method.
The total geodesic distances $\lambda_i$ between two adjacent layers $i-1$ and $i$ is defined as 
\begin{equation}
\lambda_i = \sqrt{ \sum_\vec{r} \phi_i^2(\vec{r}) } \,\,,
\end{equation}
where $\phi_i(\vec{r})$ is the rotation angle (or \textit{geodesic distance}) of the spin $\vec{n}(\vec{r})$ at site $\vec{r}$ from layer $i-1$ to $i$.
The reaction coordinate $\Lambda(i)$ is the sum of all distances $\lambda_n$, $n<i$, and serves as the measure for the evolution of the texture in Figs.~\ref{fig1},\ref{fig7},\ref{fig9}.
Following Ref.~\citenum{bessarab2015method}, the spring forces between adjacent images are proportional to the distance $\lambda_i$.
However, for the calculation of some data sets we decided to modify this part of the method by considering the actual length of the nudged band instead.
A cubic interpolation of the band in the combined space of energy and reaction coordinate is already implemented as a part of the method and it is used for plotting the continuous interpolations in Figs.~\ref{fig1},\ref{fig7},\ref{fig9}.
By using a numerical approximation of the length of this curve we achieve an improved distribution of images along the nudged elastic band and therefore observe an improved convergence behavior.
A calculation of a minimal energy path is converged when the slope of the elastic band in the climbing image vanishes.
In our simulations this is assumed when the total projected force on the climbing image is below $10^{-10}J$ or $10^{-10}J_1$, respectively.

%----------------------------------------------------------------------------------------
\subsection{Methods: High-accuracy numerical calculations of the continuum limit}
\label{sec:appendix:continuumlimit}
%----------------------------------------------------------------------------------------

We have shown that 4th order gradient terms play an important role for the topological energy barrier of 2d magnetic skyrmions. For the discretization of a continuum model, it is therefore essential to properly model higher-order terms. A second motivation for an improved discretization of continuum models is an improvement of numerical precision. For example, the calculation of the leading $\mathcal{O}(q^4)$ perturbative corrections to the $\mathcal{O}(q^2)$ continuum model in Sec.~\ref{sec:dmi:largeskyrmions:skyrmionenergy} builds on the exact skyrmion solution.
However, an analytic expression of this solution is not known.
We therefore calculate numerically with very high precision the skyrmion texture which minimizes the energy of the dimensionless continuum model, Eq.~\eqref{eq:dmi:model:continuous:rescaled}.

Usually, this is achieved by discretizing the continuum model on a square or triangular lattice with lattice constant $a$ and replacing derivatives with nearest neighbors finite difference approximations.
This standard approach, however, yields exactly the same formalism as the atomistic model and therefore the same $\mathcal{O}(a^2)$ scaling of the numerical error.
Therefore, when we try to calculate the texture with high numerical precision, the runtime of the simulation soon becomes critical: only a small factor $4$ in accuracy requires twice as many simulated spins per spatial direction, resulting in a factor $4$ longer computational time per iteration step. 
In addition, if we assume a constant step-width $\delta t$ for the relaxation algorithm, it is limited by $\delta t /a^2 \leq c$.
We here use $\delta t /a^2=0.1$ for most calculations.
The total runtime for the same level of convergence is therefore enhanced by a factor $16$.

\begin{table}[h!]
    \begin{tabular*}{0.48 \textwidth}{@{\extracolsep{\fill}} l| c| *{5}{c} }
        $\mathcal{O}(a^{2N})$ &$d_1$ &$w_{1,0}$ &$w_{1,\pm1}$ &$w_{1,\pm2}$ &$w_{1,\pm3}$ &$w_{1,\pm4}$ \\ \hline\hline
        $N=1$                 &$2$   &$0$   &$\pm1$     &           &           &           \\ \hline
        $N=2$                 &$12$  &$0$   &$\pm8$     &$\mp1$     &           &           \\ \hline
        $N=3$                 &$60$  &$0$   &$\pm45$    &$\mp9$     &$\pm1$     &           \\ \hline
        $N=4$                 &$840$ &$0$   &$\pm672$   &$\mp168$   &$\pm32$    &$\mp3$     \\ \hline
    \end{tabular*} 
    \caption{
        Coefficients of the higher order discretization scheme for a derivative $\partial_\alpha \vec{n}$ as defined in Eq.~\eqref{eq:appendix:continuumlimit:derivative}.
    }
    \label{tab:appendix:continuumlimit:derivative}
\end{table}
\begin{table}[h!]
    \begin{tabular*}{0.48 \textwidth}{@{\extracolsep{\fill}} l| c | *{5}{c} }
        $\mathcal{O}(a^{2N})$ &$d_2$  &$w_{2,0}$    &$w_{2,\pm1}$ &$w_{2,\pm2}$ &$w_{2,\pm3}$ &$w_{2,\pm4}$ \\ \hline\hline
        $N=1$                 &$1$    &$-2$     &$1$        &           &           &           \\ \hline
        $N=2$                 &$12$   &$-30$    &$16$       &$-1$       &           &           \\ \hline
        $N=3$                 &$180$  &$-490$   &$270$      &$-27$      &$2$        &           \\ \hline
        $N=4$                 &$5040$ &$-14350$ &$8064$     &$-1008$    &$128$      &$-9$       \\ \hline
    \end{tabular*} 
    \caption{
        Coefficients of the higher order discretization scheme for a second derivative $\partial_\alpha^2 \vec{n}$ as defined in Eq.~\eqref{eq:appendix:continuumlimit:laplacian}.
    }
    \label{tab:appendix:continuumlimit:laplacian}
\end{table}
\begin{table}[h!]
    \begin{tabular*}{0.48 \textwidth}{@{\extracolsep{\fill}} l| c | *{6}{c} }
        $\mathcal{O}(a^{2N'})$ &$d_2$  &$w_{2,0}$    &$w_{2,\pm1}$ &$w_{2,\pm2}$ &$w_{2,\pm3}$ &$w_{2,\pm4}$ &$w_{2,\pm5}$ \\ \hline\hline
        $N=2$                 &$2$    &$0$      &$\mp2$     &$\pm1$     &           &          & \\ \hline
        $N=3$                 &$8$    &$0$      &$\mp13$    &$\pm8$     &$\mp1$     &          & \\ \hline
        $N=4$                 &$240$  &$0$      &$\mp488$   &$\pm338$   &$\mp72$    &$\pm7$    & \\ \hline
        $N=5$                 &$30240$&$0$      &$\mp70098$ &$\pm52428$ &$\mp14607$ &$\pm2522$ &$\mp205$ \\ \hline
    \end{tabular*} 
    \caption{
        Coefficients of the higher order discretization scheme for a third derivative $\partial_\alpha^3 \vec{n}$ as defined in Eq.~\eqref{eq:appendix:continuumlimit:higher} with $N'=N-1$.
    }
    \label{tab:appendix:continuumlimit:three}
\end{table}
\begin{table}[h!]
    \begin{tabular*}{0.48 \textwidth}{@{\extracolsep{\fill}} l| c | *{6}{c} }
        $\mathcal{O}(a^{2N'})$ &$d_2$  &$w_{2,0}$    &$w_{2,\pm1}$ &$w_{2,\pm2}$ &$w_{2,\pm3}$ &$w_{2,\pm4}$ &$w_{2,\pm5}$ \\ \hline\hline
        $N=2$                 &$1$    &$6$      &$-4$       &$1$        &           &          &      \\ \hline
        $N=3$                 &$6$    &$56$     &$-39$      &$12$       &$-1$       &          &      \\ \hline
        $N=4$                 &$240$  &$2730$   &$-1952$    &$676$      &$-96$      &$7$       &      \\ \hline
        $N=5$                 &$15120$&$193654$ &$-140196$  &$52428$    &$-9738$    &$1261$    &$-82$ \\ \hline
    \end{tabular*} 
    \caption{
        Coefficients of the higher order discretization scheme for a 4th derivative $\partial_\alpha^4 \vec{n}$ as defined in Eq.~\eqref{eq:appendix:continuumlimit:higher} with $N'=N-1$.
    }
    \label{tab:appendix:continuumlimit:four}
\end{table}

In order to achieve faster convergence, we discretize the continuous model on a square lattice with lattice spacing $a$.
We replace simple derivatives with higher order stencils which take into account $N$ neighbors on every side:
\begin{equation}
    \partial_\alpha \vec{n} \approx \frac{1}{d_1 a} \sum_{n=-N}^N w_{1,n} \, \vec{n}(\vec{r} + n a \hat{e}_\alpha) + \mathcal{O}(a^{2N})
\label{eq:appendix:continuumlimit:derivative}
\end{equation}
where $\alpha=x,y$ is the spatial direction.
The second order derivatives are substituted by 
\begin{equation}
    \partial_\alpha^2 \vec{n} \approx \frac{1}{d_2 a^2} \sum_{n=-N}^N w_{2,n} \, \vec{n}(\vec{r} + n a \hat{e}_\alpha)  + \mathcal{O}(a^{2N}) \,\,.
\label{eq:appendix:continuumlimit:laplacian}
\end{equation}
For the discretization of the 4th order correction terms, see Sec.~\ref{sec:dmi:largeskyrmions:skyrmionenergy}, we only need the square of second derivatives.
If, however, one wishes to calculate the dynamics of this system, e.g., in order to relax the magnetic texture, then also higher order derivatives and their discretizations become inevitable.
In order to achieve the same scaling of the numerical error, these higher order contributions require more neighbors to be included:
\begin{equation}
    \partial_\alpha^{3/4} \vec{n} \approx \frac{1}{d_2 a^{3/4}} \sum_{n=-N}^N w_{3/4,n} \, \vec{n}(\vec{r} + n a \hat{e}_\alpha)  + \mathcal{O}(a^{2N-2}) \,\,.
\label{eq:appendix:continuumlimit:higher}
\end{equation}
The weights $w_{i,j}$ and $w_{i,j}$ and the denominators $d_i$ are given in Tabs.~\ref{tab:appendix:continuumlimit:derivative}, \ref{tab:appendix:continuumlimit:laplacian},\ref{tab:appendix:continuumlimit:three}, and\ref{tab:appendix:continuumlimit:four} respectively.
Furthermore, a discussion for open boundary conditions and the coefficients for the micromagnetic model including DMI can be found in Ref.~\citenum{muller2018thesis}.

This approximation scheme yields an improved $\mathcal{O}(a^{2N})$ convergence while the total runtime for a fixed number of spins only increases by a factor $N$. 
Here, we use a discretization of $a=0.05$ on a square lattice which, following Ref.~\citenum{muller2018thesis}, leads to a total error in the energy of the skyrmion of the order of $10^{-10}J$ while the standard $\mathcal{O}(a^2)$ scheme yields an error or the order $10^{-2}J$ for the same discretization.
Note that the numerical error of the $\mathcal{O}(a^2)$ scheme is, in fact, the size-dependent energy of the atomistic model, which is also derived as a result of the perturbative analysis in Sec.~\ref{sec:dmi:largeskyrmions:skyrmionenergy}.

%----------------------------------------------------------------------------------------
\section{Asymmetric decay of skyrmions}
\label{sec:appendix:asymmetricdecay}
%----------------------------------------------------------------------------------------
 
\begin{figure}
    \center
    \includegraphics[width=0.47 \textwidth]{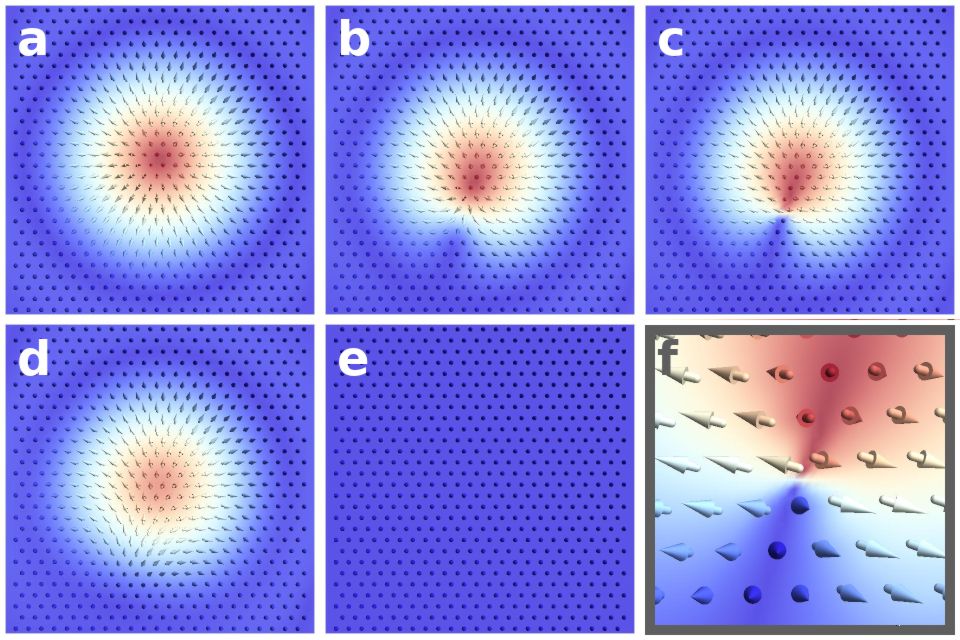}
    \vspace{1mm}

    \includegraphics[width=0.47 \textwidth]{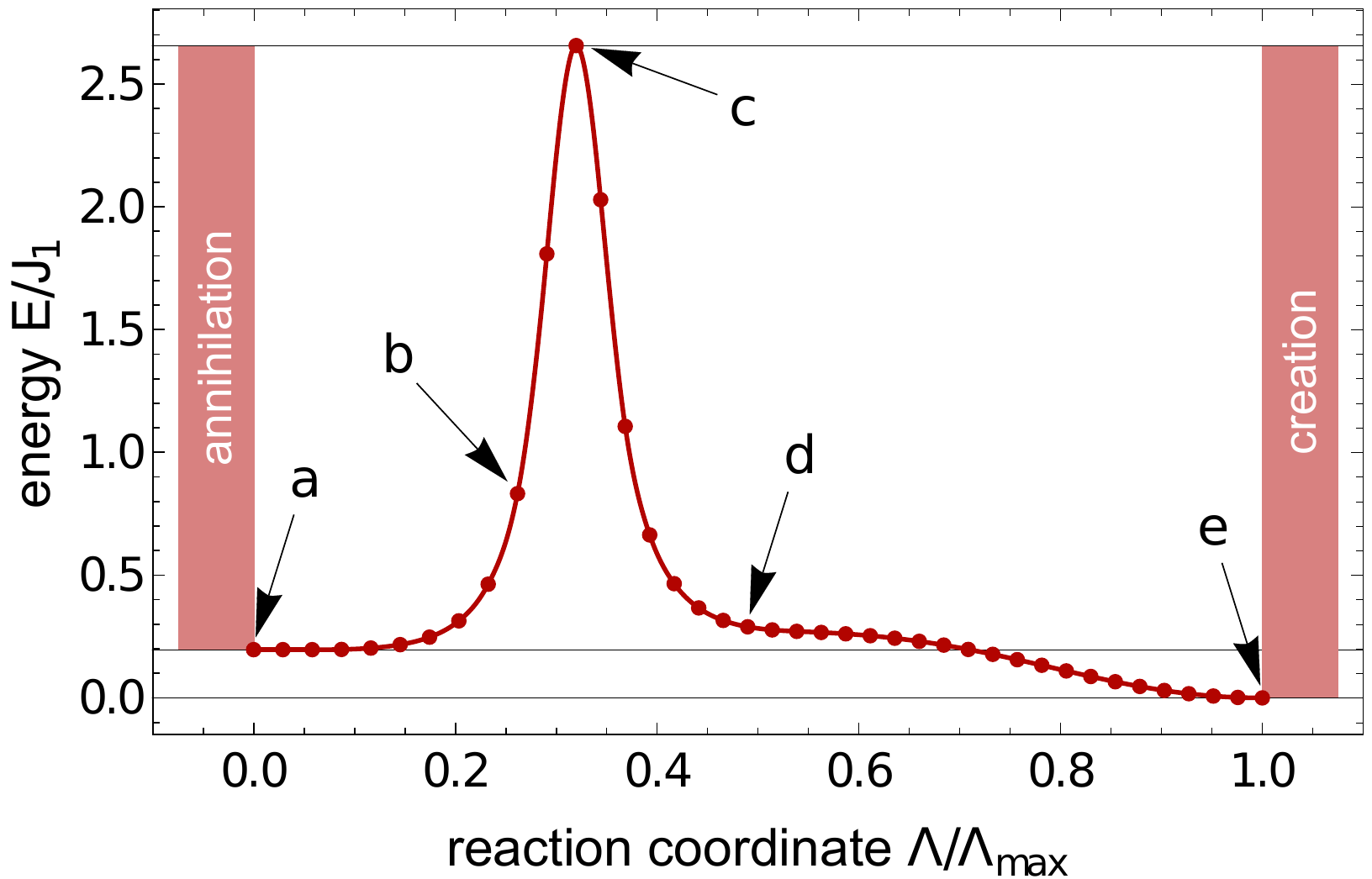}
    \caption{
        Minimal energy path for the creation/annihilation of a skyrmion in the symmetric model system, Eq.~\eqref{eq:frustration:model}, as obtained from the GNEB method, see Sec.~\ref{sec:appendix:minimalenergypath}.
        The upper panels (a-f) show the real-space magnetic texture in the proximity of the center of the skyrmion for various states along the minimal energy path.
        The color encodes the out-of-plane component of the magnetization.
        Panel (f) is a close-up of the saddle point texture in panel (c).
        The lower panel shows the energy $E/J_1$ along the minimal energy path as a function of the reaction coordinate $\Lambda$, see Sec.~\ref{sec:appendix:minimalenergypath}. 
        The energy is evaluated with respect to the polarized phase (e).
        The arrows indicate the position of the upper panels in the minimal energy path.
        These results were obtained for $J_2/J_1=31/90$, $\mu_s H/J_1=1/1800$, and $K/J_1=1/360$. 
        % system 6, scaleindex si=4
        % see Mathematica-Notebook for more details.
    }
    \label{figS2}
\end{figure}

For skyrmions in a symmetric magnet without Dzyaloshinskii-Moriya interaction we observe that the initially symmetric collapse path becomes asymmetric for certain parameter regions $(h,\kappa)$, see Fig.~\ref{figS2}, similar to the decay reported in Ref.~\citenum{meyer2019isolated}.
Both the symmetric, Fig.~\ref{fig7}, and the asymmetric collapse path have in common that they let not shrink the skyrmion radius to zero and therefore require a vortex singularity. 
In large skyrmions where the energy is dominated by the exchange interactions, the vortex core consists of three neighboring spins lying in a common plane with a $120^\circ$ angle relative to each other with subleading corrections due to field and anisotropy.
In the symmetric scenario this vortex core is depicted in Fig.~\ref{fig7}f where the common plane is the xy-plane.
In the asymmetric scenario one of the spins in the vortex core points in the $\hat{z}$-direction.

The total magnetization of a vortex core is independent of the orientation of the common plane of spins, therefore the external magnetic field $h$ does not favor any particular direction.
The uni-axial anisotropy $\kappa$, however, is proportional to $m_z^2$ and therefore favors vortex cores where one of the spins is aligned out of plane.
Therefore we only observe the asymmetric decay for the systems with a relatively large $\kappa = \frac{35}{160}$.
Note that, furthermore, the discussion in the limit of large skyrmions, Sec.~\ref{sec:frustration:largeskyrmions}, is also valid for the asymmetric decay as the energy contributions for the saddle point only arise from the vortex which is in the limit $\xi\to\infty$ not only independent of the external field and anisotropy but also of the orientation of the vortex in the plane.
Furthermore this implies that in this extreme limit the saddle point is not uniquely symmetric or asymmetric but can be any which has a negative impact on the convergence of large systems.

\end{document}